\documentclass{sig-alternate}

\pdfoutput=1

\usepackage{fancyhdr}
\usepackage[normalem]{ulem}
\usepackage[hyphens]{url}
\usepackage{amsmath}
\usepackage{multirow}
\usepackage{subfig}
\usepackage{caption}
\usepackage{float}
\usepackage[hyphens]{url}
\usepackage{amssymb}
\usepackage{mathtools}
\usepackage{xspace}
\usepackage{color}
\usepackage{comment}
\usepackage[flushleft]{threeparttable}
\usepackage[ruled,vlined]{algorithm2e}
\usepackage[noend]{algcompatible}
\usepackage{cite}
\usepackage{epstopdf}
\usepackage{pifont}
\usepackage{listings}
\usepackage{tikz}
\usepackage{enumitem}

\newcommand{\bheading}[1]{{\vspace{2pt}\noindent{\textbf{#1}}\hspace{2pt}}}

\newcommand{\RNum}[1]{\uppercase\expandafter{\romannumeral #1\relax}}

\newcommand{\etc}{\emph{etc.}\xspace}
\newcommand{\ie}{\emph{i.e.}\xspace}
\newcommand{\eg}{\emph{e.g.}\xspace}

\newcommand{\secref}[1]{\mbox{Sec.~\ref{#1}}\xspace}

\def\authnotes{1}
\newcommand{\authnote}[2]{\ifnum\authnotes=1\begin{quote}\textbf{#1 says:} #2\end{quote}\fi}
\newcommand{\fixme}[1]{\ifnum\authnotes=1\textbf{\textcolor{red}{[FIXME: #1]}}\fi}

\newenvironment{packeditemize}{
\begin{list}{$\bullet$}{
\setlength{\labelwidth}{8pt}
\setlength{\itemsep}{0pt}
\setlength{\leftmargin}{\labelwidth}
\addtolength{\leftmargin}{\labelsep}
\setlength{\parindent}{0pt}
\setlength{\listparindent}{\parindent}
\setlength{\parsep}{0pt}
\setlength{\topsep}{3pt}}}{\end{list}}

\makeatletter
\def\@copyrightspace{\relax}
\makeatother

\begin{document}
\title{Privacy-preserving Machine Learning \\through Data Obfuscation}

\numberofauthors{3}
\author{
\alignauthor
Tianwei Zhang\\
       \affaddr{Princeton University}\\
       \email{\normalsize{tianweiz@alumni.princeton.edu}}
\alignauthor
Zecheng He\\
       \affaddr{Princeton University}\\
       \email{\normalsize{zechengh@princeton.edu}}
\alignauthor
Ruby B. Lee\\
       \affaddr{Princeton University}\\
       \email{\normalsize{rblee@princeton.edu}}
}
\maketitle

\begin{abstract}
As machine learning becomes a practice and commodity, numerous cloud-based services and frameworks are provided to help customers develop and deploy machine learning applications. While it is prevalent to outsource model training and serving tasks in the cloud, it is important to protect the privacy of sensitive samples in the training dataset and prevent information leakage to untrusted third parties. Past work have shown that a malicious machine learning service provider or end user can easily extract critical information about the training samples,  from the model parameters or even just model outputs. 

In this paper, we propose a novel and generic methodology to preserve the privacy of training data in machine learning applications. Specifically we introduce a \textsc{obfuscate} function and apply it to the training data before feeding them to the model training task. This function adds random noise to existing samples, or augments the dataset with new samples. By doing so sensitive information about the properties of individual samples, or statistical properties of a group of samples, is hidden. Meanwhile the model trained from the obfuscated dataset can still achieve high accuracy. With this approach, the customers can safely disclose the data or models to third-party providers or end users without the need to worry about data privacy. Our experiments show that this approach can effective defeat four existing types of machine learning privacy attacks at negligible accuracy cost. 

\end{abstract}

\section{Introduction}
\label{sec:intro}

Nowadays machine learning has become a core component of many popular applications, such as image classification, speech recognition, natural language translation. To support the increased demand of machine learning-based development, many cloud providers push Machine Learning as a Service (MLaaS), providing computing platforms and learning frameworks for machine learning model training and serving. Typical MLaaS platforms include Amazon Sagemaker \cite{amazonsagemaker}, Google Cloud ML Engine \cite{googleMLEngine}, Microsoft Azure ML Studio \cite{microsoftazure}. To use the MLaaS, customers provide training dataset and (or) machine learning algorithms to the cloud provider. The cloud provider sets up the machine learning environment, allocates certain amount of computation resources, and runs the model training task automatically. The cloud provider can also offer the model serving services, in which a model trained in the cloud side or the customer side is stored in the cloud platform. The cloud provider releases query APIs to end users, enabling them to use the model for prediction or classification. 

A training dataset is necessary to train and generate a machine learning model. A dataset can contain sensitive samples, \eg, personal medical records, employee information, financial data, \etc How to protect the privacy of such samples has become a new security concern in machine learning. This privacy threat is especially serious in Machine Learning as a Service. First, in a model training service, customers need to upload the training data to the cloud provider, and the provider has full access to the data. Past work have shown that a malicious provider can easily steal the sensitive data, embed them into the model for leakage \cite{SoRiSh:17}. Second, in a model serving service, customers need to upload the pre-trained models to the cloud provider and set up end points for remote users to use the models. A successful machine learning model incorporates essential information about the training set. So even if the malicious provider does not have direct access to the dataset, he can extract sensitive information about the training data from the model parameters \cite{AtMaSp:15, HiAtPe:17, MeSoDe:18}. Third, even if the cloud provider is trusted, a malicious remote user who has only black-box access to the model output results is still able to retrieve information about the training data by querying the model with carefully-crafted inputs \cite{FrLaJh:14, FrJhRi:15, ShStSo:17, JaLuGe:17, LiLiGa:18, SaZhHu:18, YeGiFr:18, LoBiWa:18}.


Protecting the privacy of training data in machine learning is challenging. First, when customers outsource machine learning tasks to the MLaaS provider, they have to grant the service provider full accesses to the data or models. This inevitably causes a big attack surface. Second, a correct machine learning model \emph{must} reveal information about the data on which it was trained. This creates opportunities for an adversary to learn information from the models. Third, given a fact that the adversary can steal different properties of the sensitive samples from different sources in different scenarios, it is very hard to have a generic method to eliminate all these information leakage. 

Some solutions were proposed to achieve privacy-preserving machine learning. These solutions can be classified into two categories. The first category is to design privacy-aware machine learning algorithms \cite{ShSh:15, AbChGo:16} and platforms \cite{OhScFo:16, HuSoSh:18} to prevent information leakage from models. However, this assumes that these security schemes can be correctly deployed in MLaaS. A malicious cloud provider with full control of machine learning tasks and platforms can easily subvert such privacy protection by disabling these solutions. Besides, in a model training service, customers are required to disclose the dataset to the cloud provider. These security-aware algorithms or platforms cannot prevent training data leakage in this scenario. The second category is to process the training data before releasing them to untrusted third parties. For instance, homomorphic encryption is introduced to process the training samples and then models are trained over the encrypted data \cite{BoPoTu:15} . While this can prevent data leakage, homomorphic encryption is not generic for arbitrary machine learning algorithms and the performance is not efficient.

In this paper, we propose a novel and generic methodology to protect the privacy of training data in MLaaS. Our method follows the second defense category described above: training data processing. Specifically, we introduce a new function \texttt{Obfuscate} to help hide the sensitive information by adding random noise to the training dataset. We consider two cases: for the attacks that attempt to steal the properties of individual samples, we add random noise to these sensitive samples; for the attacks that attempt to learn the statistical properties of groups of samples, we augment the dataset with faked noisy samples. These noise are injected to obfuscate the adversary's observations, while maintaining the prediction accuracy of trained models. Customers who are concerned with their sensitive data can apply this function to the training dataset and then disclose them to the MLaaS provider. So even the untrusted service provider or remote end users have access to the training data, model parameters or model outputs, they cannot discover the sensitive information hidden by the \texttt{Obfuscate} operation. Our evaluation indicates that this methodology can eliminate the leakage of four different types of privacy attacks \cite{FrJhRi:15, AtMaSp:15, ShStSo:17, SoRiSh:17}. This function introduces negligible performance overhead and accuracy reduction ($<5\%$).

The key contributions of this paper are:

\begin{packeditemize}

\item a comprehensive analysis of privacy attacks in machine learning services.

\item a generic method of obfuscating training dataset to eliminate data leakage.

\item evaluation of data obfuscation to defeat different types of machine learning privacy attacks. 

\end{packeditemize}

The rest of the paper is organized as follows: \secref{sec:bg} gives the background of machine learning and model privacy attacks. \secref{sec:method} describes the methodology of obfuscating individual or group of samples to hide information. \secref{sec:case} shows the application of this methodology in different case studies. We discuss related work in \secref{sec:related} and conclude in \secref{sec:conclu}.

\section{Background}
\label{sec:bg}

\subsection{Machine Learning}
\label{sec:bg:dl}

A supervised machine learning model is a parameterized function $f_\theta: \mathcal{X} \mapsto \mathcal{Y}$ that maps input data $x\in \mathcal{X}$ (feature) to output data $y\in \mathcal{Y}$ (label). For a classification problem, $\mathcal{X}$ is a $d$-dimensional vector space and $\mathcal{Y}$ is a discrete set of classes. This function is able to accurately predict the label of new data that have not seen before. 

\bheading{Model training.} The training process of a machine learning model is to find the optimal parameters $\theta$ that can accurately reflect the relationship between $\mathcal{X}$ and $\mathcal{Y}$. To achieve this, a training dataset $D=\{x_i, y_i\}^{N}_{i=1}$ with $N$ samples is needed, where $x_i \in \mathcal{X}$ is the feature data and $y_i \in \mathcal{Y}$ is the corresponding ground-truth label. Then a loss function $L$ is adopted to measure the distance between the ground-truth output $y_i$ and the predicted output $f_\theta(x_i)$. $\Omega$ is a regularization term that penalizes model complexity and help avoid overfitting. The goal of training a model is to minimize this loss function (Equation \ref{eq:loss1}).

\begin{align}
\label{eq:loss1}
\theta^* = \arg\min_{\theta}(\sum_{i=1}^N L(y_i, f_\theta(x_i) + \Omega(\theta))
\end{align}

\bheading{Model inference.} After the model training is completed and the optimal parameters $\theta^{*}$ are obtained, given an input $x$, the corresponding output can be calculated as $y = f_{\theta^*}(x)$. This prediction process is called inference. We can also calculate the prediction accuracy of the model over a testing dataset $D^{t}=\{x^{t}_i, y^{t}_i\}^{M}_{i=1}$ to measure the model's performance. For a classification problem where the output is a discrete number of values, the prediction accuracy is defined in Equation \ref{eq:accuracy}, where $\mathbb{I}$ is the indicator function.

\begin{align}
\label{eq:accuracy}
\begin{split}
acc(D^{t}, f_{\theta^*}) = \frac{1}{M}\sum_{i=1}^{M}\mathbb{I}(f_{\theta^*}(x^{t}_i) = y^{t}_i)
\end{split}
\end{align}

\bheading{Deep learning.} The past few years have witnessed the fast development in deep learning, a new branch of machine learning. Algorithmic breakthroughs, the feasibility of data collection, and increasing computational power have contributed to the popularity of deep learning techniques. Deep learning have been widely adopted in many artificial intellectual applications, such as image recognition \cite{HeZhRe:15}, natural language processing \cite{LuPhMa:15}, speech recognition \cite{HaCaCa:14}, \etc. 

A deep learning model is usually in the form of a neural network. Various neural network architectures have been proposed for different applications, \eg multilayer perceptrons \cite{Ro:58}, convolutional neural networks \cite{LeJaBo:89} and recurrent neural networks \cite{RuHiWi:86}. A neural network consists of an input layer, an output layer and a sequence of hidden layers between the input and output. Each layer is a collection of units called \emph{neurons}, which are connected to other neurons in the previous layer and the following layer. Each connection between the neurons can transmit a signal to another neuron in the next layer by applying a linear function followed by an element-wise nonlinear activation function (e.g. sigmoid or ReLU). In this way, a neural network transforms the inputs through hidden layers and then the outputs. Backward propagation \cite{GoBeCo:16} and stochastic gradient descent \cite{RoMo:51} are commonly used methods to train a deep learning model and find the optimal parameters.

\subsection{Privacy Threats in MLaaS}
\label{sec:bg:threat}

Cloud-based machine learning services are released to provide automated solutions for machine learning data processing, model training, inference and further deployment. They attract machine learning practitioners to deploy applications in the cloud without having to set up their own large-scale infrastructure and computation resources. 

\begin{figure}[t]
     \centering
     \subfloat[][Model training]{
     \includegraphics[width=0.38\linewidth]{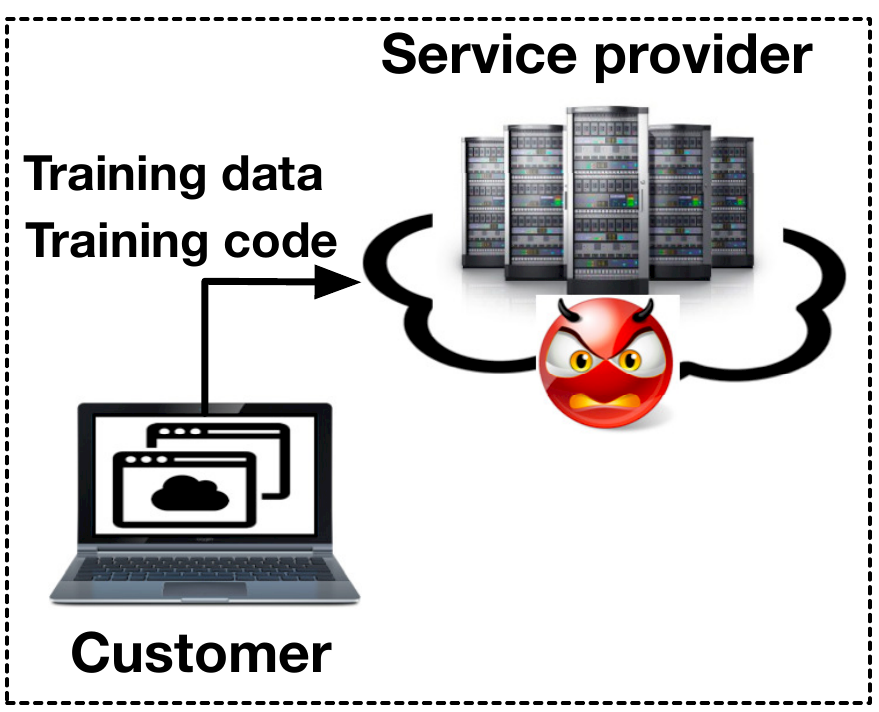}
     \label{fig:model_training}} \hspace{1em}%
     \subfloat[][Model serving]{
     \includegraphics[width=0.53\linewidth]{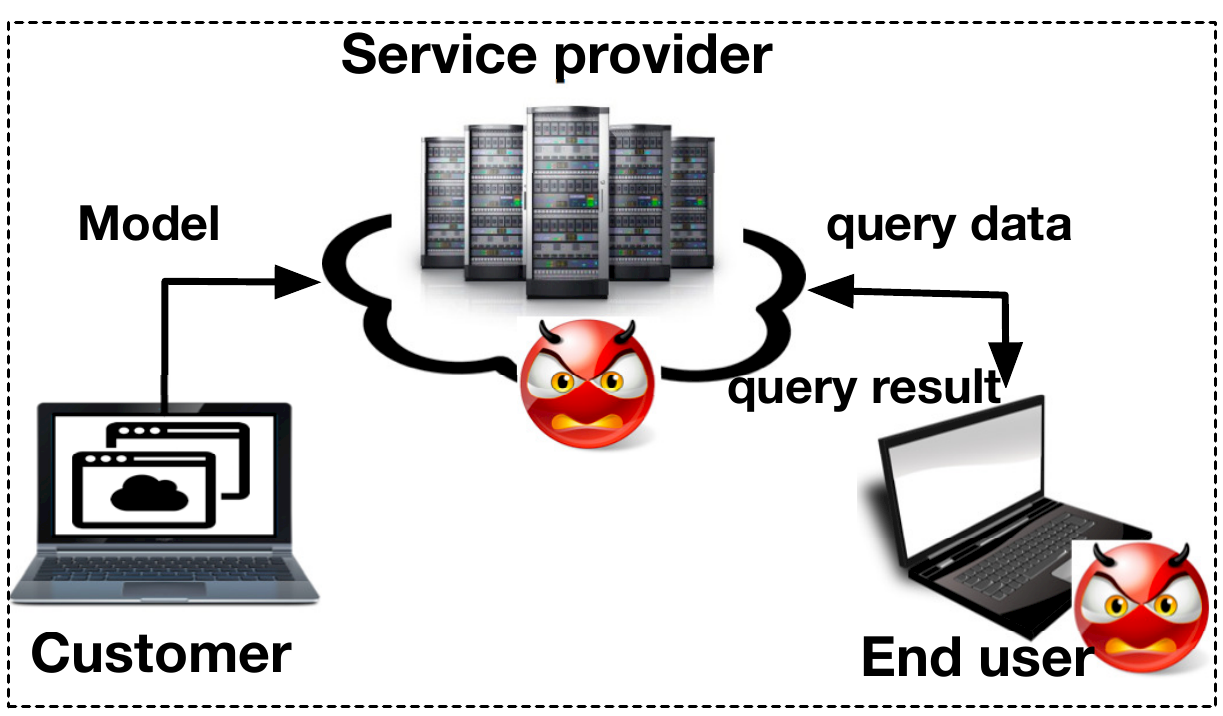}
     \label{fig:model_serving}}
    \caption{Data leakage in MLaaS.}
    \label{fig:data_leakage}
\end{figure}

There are two main types of machine learning services in the cloud (Figure \ref{fig:data_leakage}). The first one is model training: the cloud provider offers the computing resources and the customer selects a machine learning algorithm. Then the ML algorithm runs on the computing resources and generates the model for the customer. The second one is model serving: the customer uploads a model to the cloud platform. The cloud platform releases query APIs to end users. Then the end users can use the ML model for prediction or classification via these query APIs. A cloud platform can offer both the model training and model serving services to customers to achieve end-to-end machine learning deployment. 

Recent work proposed privacy attacks in MLaaS: a malicious cloud provider or end user can extract sensitive information about the training dataset from machine learning tasks running in the cloud. We summarize these attacks based on the leaked information sources and types.

\subsubsection{Leaked information sources.}

In different service modes, an adversary can steal sensitive information from different sources. 

\bheading{Training set.}
Generating a model requires the training data and code. In the model training service (Figure \ref{fig:model_training}), customers do not necessarily need to upload the training code as they can select default ones from the cloud provider. However, training data is a necessity from customers to generate the models that are applicable to customers' scenarios. So the cloud provider has full access to the training data $D$. If the cloud provider is malicious, he can retrieve arbitrary information directly from the training set. Song et al. \cite{SoRiSh:17} designed a ML model memorization attack based on this threat model: the adversary can steal the critical training data and encode them into the trained model in different ways. So another adversary is able to retrieve the training data from the model parameters if he has white-box access to the model, or from the model outputs if he only has black-box access to the model. 

\bheading{Model parameters.}
In the model serving service (Figure \ref{fig:model_serving}), the cloud provider may not have access to the training set if the model is not trained in the cloud. However he has access to the trained model $f_\theta$ and is able to identify the model algorithm, neural network topology, model parameters and hyper-parameters. Since the model $f_\theta$ is generated over the training $D$ (Equation \ref{eq:loss1}), the model parameters $\theta$ reveal some information about the training set. A malicious cloud provider is able to steal some information from the model parameters. For instance, Ateniese et al. \cite{AtMaSp:15} demonstrated a machine learning attack, that uses model parameters as features to classify and identify some sensitive properties about the training dataset. 

In the model training service, model parameters can also leak information. This case happens in distributed machine learning, where multiple nodes collaborate to train a machine learning model and each node uses its own training dataset. As every node needs to keep synchronizing the model parameters with other nodes. A malicious node can infer the training dataset of other nodes from the shared model parameters. Membership inference attack \cite{MeSoDe:18} and model inversion attack \cite{HiAtPe:17} have been demonstrated in this setting and threat model.

\bheading{Model output.}
In the model serving scenario (Figure \ref{fig:model_serving}), in addition to the cloud provider who has white-box access to the model, an end user with black-box access to the model also has chances to retrieve training set properties only from the model outputs. Basically end users are allowed to use the model for prediction or classification. They send a sample to the cloud provider and receive the corresponding prediction result for this sample. The idea here is that the training data can affect the model parameters, which can then affect the prediction results of specific samples. Then an untrusted end user can send carefully-crafted input to the model and use the output to learn about the training data. For instance, in model inversion attacks \cite{FrLaJh:14, FrJhRi:15}, the adversary attempts to reconstruct the training sample from the output confidence score. The attack strategy is to find the optimized sample whose confidence score is most close to 1 for one specific class. In membership inference attacks \cite{ShStSo:17}, the adversary uses the output confidence score of some samples as the training data to build a classify and predict if one target training sample is a member of the training set used for training this model.

\subsubsection{Leaked information types.}
\label{sec:leaked_type}
The adversary can extract different types of properties about the training set $D=\{x_i, y_i\}^{N}_{i=1}$, as described below.

\bheading{Individual samples.}
An adversary may be interested in some specific training samples $(x_{i^*}, y_{i^*})$ inside the training set, and want to obtain the properties of these samples, denoted as $\mathcal{P}(x_{i^*}, y_{i^*})$. 

For instance, the adversary wants to fully recover the sensitive training samples. In this case, the sensitive property is the sample values, \ie, $\mathcal{P}(x_{i^*}, y_{i^*}) = x_{i^*}$. Past work have shown that the adversary is able to steal the sensitive medical records in pharmacogenetic models via model inversion attacks \cite{FrLaJh:14} or images in image classification models \cite{SoRiSh:17}. 

Another example is membership inference attack, where the adversary wants to infer if one sample is used during the model training phase. Then the property is denoted as $\mathcal{P}(x_{i^*}, y_{i^*}) = \mathbb{I}(x_{i^*}, y_{i^*} \in D)$. Shokri et al. \cite{ShStSo:17} demonstrated the possibility of membership inference attacks in different machine learning models, \eg, image classification, purchase recommendation, medical procedure prediction.


\bheading{Groups of samples.}
Instead of one individual sample, the adversary may be interested in a group of samples $g^{*}$. In this group, each sample $(x_{i}, y_{i}) \in g^{*}$ has the same feature $\mathcal{F}$. The adversary wants to learn the statistical properties of these samples, denoted as $\mathcal{P}(g^{*})$.

One typical example is model inversion attack \cite{FrJhRi:15}. In a face recognition system, the adversary wants to reconstruct the looking of one person (\ie, one class). In this case, the adversary considers a group of samples within the same class: $\mathcal{F}: y_{i} = y^{*}$. The property he wants to obtain is the average value of the samples in this group: $\mathcal{P}(g^{*}) = avg (x_{i})$. This attack is effective at the datasets that samples within each class is very similar.

A similar attack is demonstrated in \cite{AtMaSp:15}, where the adversary can extract the nationalities of samples in the speech recognition system, and the traffic source website in the traffic analyzer. Here the adversary considers all the samples in the dataset: $\mathcal{F}: x_{i}, y_{i} \in D$, \ie, $g^{*} = D$. The properties in consideration are some attributes of the samples, $\mathcal{P}(g^{*}) = attr(x_i)$. These properties can be recovered from model parameters.

\section{Methodology}
\label{sec:method}

\bheading{Threat model.}
We consider three attack settings in two machine learning service modes, as shown in Figure \ref{fig:data_leakage}. In the first case, the adversary is the cloud provider in the model training mode (Figure \ref{fig:model_training}). The customers upload the training data to the adversary. So the adversary has direct access to all the training data. In the second case, the adversary is also the cloud provider in the model serving mode (Figure \ref{fig:model_serving}). The customers upload the model to the adversary. So he has \emph{white-box} access to the target model: \ie, he is able to access the model parameters (e.g., network weights), hyper-parameters (e.g., regularization parameters, number of training epochs), model architecture. In the third case, the adversary is a remote end user in the model serving mode (Figure \ref{fig:model_serving}). He only has \emph{black-box} access to the model: \ie, he can only make queries to the target model and get the prediction results, \eg, the confident score of each class for a given input in a classification model.

\bheading{Challenges.}
It is very challenging to design a method to protect the privacy of training data in MLaaS. First, the features of ML and MLaaS make it hard to eliminate information leakage to third parties. For model training, the customers \emph{must} disclose the data to the cloud provider to run the training task. For model serving, the customers \emph{must} allow the cloud provider to access the model parameters, and the end users to obtain the model outputs. The parameters and outputs of a correct ML model \emph{inevitably} reflect the training data. 

Second, the adversary in our threat model is very powerful and it is difficult to have a generic method to cover all attack settings. Past work proposed new algorithms or systems to preserve the training data privacy \cite{ShSh:15, OhScFo:16, AbChGo:16}. These work all assume that the desired algorithms and systems are correctly implemented and deployed in MLaaS. For the threat model in which the adversary has full control of the training dataset and model training process, enhancing the training algorithms or environment is not enough to guarantee the privacy of training data. 

Third, there are different types of properties about the training set (\secref{sec:leaked_type}), and considering all theses properties is challenging. For instance, collaborative machine learning with differential privacy was proposed to protect the privacy of training data \cite{ShSh:15}. However, it has been proven that this method can only protect the properties of individual samples, but not the statistical properties of groups of samples \cite{HiAtPe:17, MeSoDe:18}. Similarly, Triastcyn and Faltings \cite{TrFa:18} proposed to use GAN to generate artificial dataset to replace the original dataset for privacy-aware model training. However, the artificial dataset still keeps the statistical properties of original dataset, as the discriminator network in GAN cannot distinguish the artificial dataset from the original one.

\bheading{Overview.}
We aim to design a methodology to protect the privacy of training dataset in MLaaS. Our goal is to eliminate the leakage of properties of both individual samples as well as groups of samples, even the adversary (malicious cloud provider or end users) can access the training dataset, model parameters or model outputs. 

Instead of designing new algorithms or systems, we propose to process the training data $D$ and convert it to $D'$ before outsourcing them into the machine learning services. This data processing needs to meet two goals:

\begin{packeditemize}

\item \textbf{R1}: there are no confidential information about critical individual or groups of samples embedded in $D'$ or the model trained from such data. So the adversary with access to $D'$, the model or the query outputs cannot get useful information.

\item \textbf{R2}: The model trained from the new dataset $D'$ should have negligible prediction accuracy decrease compared to the one trained from the original dataset $D$.

\end{packeditemize}

\begin{figure}[t]
\centerline{\mbox{\includegraphics[width=\linewidth]{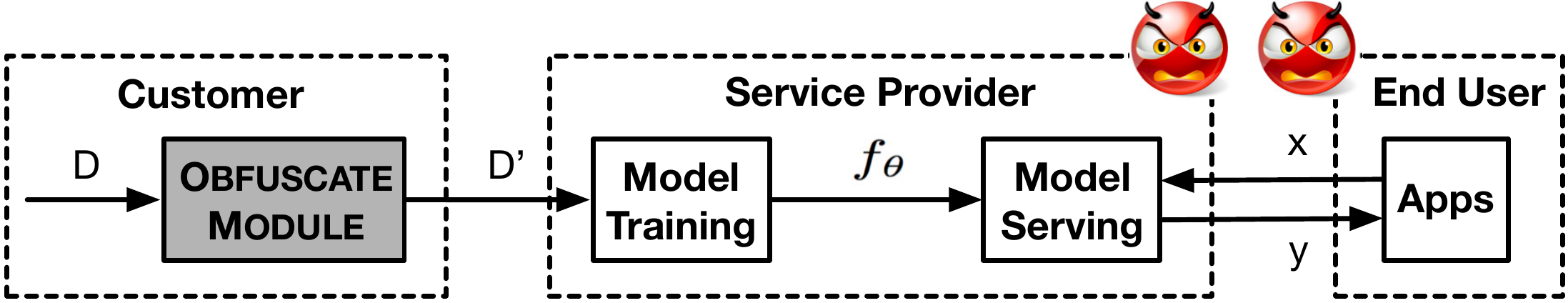}}}
\caption{Methodology Overview}
\label{fig:method:overview}
\end{figure}

Figure \ref{fig:method:overview} shows the overview of our methodology. We add a new module, \textsc{Obfuscate Module} at the customer's side to process the training data. This module obfuscates the training dataset, to hide critical information. With this design we do not need to make any assumptions or requirements for the untrusted third parties. We consider the privacy of individual properties as well as group statistical properties. To hide individual properties, we add random noise to sensitive samples. To hide group statistical properties, we augment the dataset with new faked samples. Below we describe the obfuscation mechanisms to achieve the two goals.


\subsection{Obfuscating Individual Samples}
\label{sec:method:individual}

\subsubsection{Mechanism}

We first consider the protection of individual samples. Given a dataset $D=\{x_i, y_i\}^{N}_{i=1}$, there are some sensitive samples that customers do not want to disclose to third parties. We denote the set of these samples as $d^{*} \subseteq D$. Note that the samples inside this set do not need to have common features. This is different from the groups of samples discussed in \secref{sec:method:group}. For each sensitive sample $(x^*, y^*) \in d^{*}$, the adversary is interested to learn the property $\mathcal{P}(x^*, y^*)$.

The \textsc{Obfuscate Module} takes these sensitive samples as input, and converts them to new samples. Algorithm \ref{alg:ob_individual_sample} shows the pseudo code of this module. The \texttt{ObfuscateIndividualSample} function inside this module is applied to obfuscate each sensitive sample. The basic idea is to add noise to these samples, so the adversary will not learn the property from the converted new samples. Meanwhile, the added noise will not affect the model accuracy. 

There are different ways to add noise. Generically, for the input vector $x_{i^*}$, we can randomly select part of the feature values (with the ratio of $r$) and add Gaussian noise $\mathcal{N}(0, \sigma)$ to these values. Requirements \textbf{R1} and \textbf{R2} can be achieved by adjusting the parameters $r$ and $\sigma$. A more sophisticated way is to identify the sensitive features in each input and only add Gaussian noise $\mathcal{N}(0, \sigma)$ to these feature values. For instance, in a face recognition model, we can only obscure the faces of sensitive training images. For a medical machine learning model, we can add noise to the critical features, \eg, age or occupation, of the sensitive patients.

\begin{algorithm}
\scriptsize
\SetAlgoLined
 \KwIn{}
 \Indp $D$: original dataset\\
       $d^{*}$: the set of sensitive samples\\
\Indm \KwOut{}
\Indp $D'$: new artificial dataset\\
\Indm
\SetKwFunction{proc}{ObfuscateIndividualSample}
\SetKwProg{myalg}{function}{}{end}
    \BlankLine
  	\BlankLine
\Begin{
  $D' = D$ \\
  \For{\emph{each} $(x^*, y^*) \in d^*$ } {   
    remove $(x^*, y^*)$ from $D'$ \\ 
    $({x^*}', {y^*}')$ = \texttt{ObfuscateIndividualSample}$(x^*, y^*)$ \\
    add $({x^*}', {y^*}')$ to $D'$ \\ 
    }
    \KwRet{$D'$} 
  } {}
    \BlankLine
  	\BlankLine

\myalg{\proc{$x^*, y^*$}}{
    ${x^*}'$ = $x^*$ + \texttt{Noise}($x^*$)\\
    ${y^*}'$ = $y^*$ \\
  \BlankLine
  \KwRet $({x^*}', {y^*}')$
}
 \caption{Obfuscating individual samples}
 \label{alg:ob_individual_sample}
\end{algorithm}

\subsubsection{Analysis}

When adding noise to individual samples, there is a trade-off between the privacy protection (\textbf{R1}) and model prediction accuracy (\textbf{R2}).

For requirement \textbf{R1}, adding random noise to individual samples can reduce the information leakage about these samples. Consider the case in which the adversary has the maximum capability, \eg, direct access to the training dataset \cite{SoRiSh:17}. The adversary can retrieve the maximum and most accurate information about the training data, \ie, the exact values. After obfuscating the sensitive samples, the adversary can at most obtain the samples with random noise. If the strength of added noise ($\sigma$) is large enough and the ratio of feature values involved ($r$) is large enough to cover sensitive information, the adversary fail to get useful information from his observation. For the threat model where the adversary does not have direct access to the dataset, he has to infer the property of the individual samples indirectly from the model parameters or query outputs \cite{ShStSo:17}. The models trained over the converted samples can only reflect the characteristics of noisy ones, and it is difficult for the adversary to recover the correct information. 

For requirement \textbf{R2}, machine learning models are robust against appropriate random noise in the training set \cite{ZhZhKa:17}. If the added noise is not biased, it can help prevent model overfitting. However, adding too much noise to the training dataset can reduce the accuracy of the target model. There are two factors that affect the model accuracy when obfuscating the samples: the number of samples that need to be obfuscated, and the amount of noise added to each sensitive sample. If only a small number of samples inside the training set need to be protected and obfuscated, then the obfuscating operation does not affect the accuracy, as the other samples can still contribute to generating a correct model. When the number of sensitive samples is large, then the amount of the noise becomes important in maintaining the model accuracy. We need to select the appropriate noise strength ($\sigma$ in Gaussian noise), and the portion of involved feature values ($r$) to hide sensitive information with negligible accuracy degradation. Experiments in \secref{sec:case} demonstrate that appropriate noise addition parameters can help achieve both requirements \textbf{R1} and \textbf{R2}.

\subsection{Obfuscating Groups of Samples}
\label{sec:method:group}

\subsubsection{Mechanism}

Next we consider the protection of statistical properties of training samples. The adversary is interested in a group of samples $g^*$ that have common feature $\mathcal{F}^*$. He expects to retrieve the statistical information about the samples in this group, $\mathcal{P}(g^*)$. The adversary may be interested in multiple groups. We denote the group of these groups as $G^*$.

Similarly, the \textsc{Obfuscate Module} uses the \texttt{ObfuscateGroupSample} function to add random noise to each sensitive group, as shown in Algorithm \ref{alg:ob_group_sample}. Different from obfuscating individual samples which directly adds noise to every sensitive sample, we can add some carefully-crafted training samples into each group instead of modifying the original samples. This operation can hide the statistical properties of sensitive groups.

We introduce a new parameter $r$, denoting the ratio of new samples added to a group in consideration. Then we calculate the number of new samples required for this group. To generate a new sample, we randomly pick an original one from this group, convert the feature value into its negative, and add small amount of noise. We assign this new feature value with a label so that this new sample satisfies the group's common feature $\mathcal{F}^*$. For instance, for a face recognition task, the negative of a pixel value $x$ of an image sample is calculated as $255-x$. If the adversary attempts to reconstruct the looking of one person, then the common feature of the target group is that all samples have the same label. So we should assign the new samples with the same label. Then we add this new sample into the group. Since the new samples are the negative of original ones, the statistical values of the new group the adversary can infer will not reveal the properties of original samples.

\begin{algorithm}
\scriptsize
\SetAlgoLined
 \KwIn{}
 \Indp $D$: original dataset\\
       $G^*$: A set of sensitive groups\\
       $r$: ratio of added samples compared to original ones \\
\Indm \KwOut{}
\Indp $D'$: new artificial dataset\\
\Indm
\SetKwFunction{proc}{\texttt{ObfuscateGroupSample}}
\SetKwProg{myalg}{function}{}{end}
  \BlankLine
  \BlankLine
\Begin{
  $D' = D$ \\
  \For{\emph{each} $g^*$ $\in$ $G^*$} {    
    remove $g^*$ from $D'$ \\ 
    ${g^*}'$ = \texttt{ObfuscateGroupSample}$(g^*, r)$ \\
    add ${g^*}'$ to $D'$ \\ 
    }
    \KwRet{$D'$} 
  }    
  \BlankLine
  \BlankLine

  \myalg{\proc{$g^*$, $r$}}{
  N = \texttt{GetNumberSample}($g^*$) $\cdot$ r \\
  ${g^*}'$ = $g^*$ \\
  \For{i = 1 ... \emph{N}} {
    get a random sample $(x^*, y^*)$ from $g^*$ \\
    ${x^*}'$ = \texttt{Negative}($x^*$) + \texttt{Noise}($x^*$) \\
    select ${y^*}'$, s.t. $({x^*}', {y^*}')$ satisfies $\mathcal{F}^*$ \\
    add $({x^*}', {y^*}')$ to ${g^*}'$ \\
  }
  \BlankLine
  \KwRet ${g^*}'$
}{}
 \caption{Obfuscating group samples}
 \label{alg:ob_group_sample}
\end{algorithm}

\subsubsection{Analysis}

Since the adversary does not focus on individual samples, obfuscating individual samples (\secref{sec:method:individual}) cannot achieve requirement \textbf{R1}: if each sample is injected with random noise, the adversary can still obtain the statistical properties of the samples, as the noise will be filtered when the adversary calculates the statistical characteristics of the samples. Instead, we add new samples whose properties are the opposite of the original ones. Then when the adversary tries to infer the properties from the models, he will get the statistical results combining both the original samples and the new samples. The injection of the new samples will hide the properties of original samples.

For requirement \textbf{R2}, adding noisy synthetic data into the training set does not affect model accuracy. It has been shown that machine learning models, especially deep learning models, have a vast capacity and they can essentially express any function to fit the given data \cite{ZhBeHa:16}. When adding new synthetic samples, the trained model from augmented dataset is fitted to both the original samples and the new ones \cite{SoRiSh:17}. So the model still has high quality on the original data samples. Experiments in \secref{sec:case} confirm this argument.

\section{Case Studies}
\label{sec:case}

In this section, we evaluate the effects of the data obfuscation in achieving privacy-preserving machine learning. We consider four existing types of model privacy attacks: model memorization attack \cite{SoRiSh:17} and membership inference attack \cite{ShStSo:17} target on the properties of individual samples; model inversion attack \cite{FrJhRi:15} and model classification attack \cite{AtMaSp:15} target on the statistical properties of groups of samples. For each attack, we demonstrate the security evaluation as well as the performance impact of obfuscating individual samples or groups of samples.

\subsection{Model Memorization Attack}

\subsubsection{Attack}
Model memorization attack targets on individual samples: the adversary is interested in some sensitive samples $(x^*, y^*)$, and wants to recover the exact feature values of these samples, \ie, $\mathcal{P}(x^*, y^*)=x^*$. The adversary can be a malicious cloud provider in a model training service, with full access to the training set and models. He can steal the sensitive samples, and encode the values into the model parameters or outputs. Another malicious party can retrieve these sensitive information from the model during model serving.

Song et al. \cite{SoRiSh:17} proposed several techniques for the adversary to encode sensitive data into the models. (1) LSB encoding: the adversary can encode the training dataset in the least significant (lower) bits of the model parameters. (2) Correlated value encoding: the adversary can gradually encode information while training model parameters. For instance, the adversary can add a malicious term to the loss function which maximizes the correlation between the parameters and the data he wants to encode. (3) Sign encoding: similar to correlated value encoding, the adversary can use the sign of model parameters to interpret as bit strings, \eg, positive parameters represent 1 and negative parameters represent 0. (4) Abusing model capacity: different from the above attacks, the adversary is assumed to have no access to the model parameters. Then he can augment the training dataset with synthetic inputs whose labels encode the critical information. Then information is leaked via the outputs of these added inputs. 

\bheading{Implementation}. We use the same code and configuration from \cite{SoRiSh:17} to reproduce the sign encoding attack, as it has a relatively good performance in data encoding. The target machine learning model is a 34-layer residual network over CIFAR10 dataset \cite{KrHi:09}. CIFAR10 dataset contains 50,000 training images from 10 classes, \eg, airplane, automobile, bird, etc. We pick one image for each class as the target sample, as shown in Figure \ref{fig:method:model_memorization_1}. Note that the images in the original CIFAR10 database are colored. Here we show them as gray-scaled as the deep learning application first converts these images to gray-scaled during input pre-proposing.

The adversary is able to encode around 400 images into one model due to the limited number of model parameters. Figure \ref{fig:method:model_memorization_2} shows the images recovered from the compromised models. We can see that the adversary can accurately recover most pixels of the images. He can easily identify the objects and details from the recovered images. 

\subsubsection{Defense}
Since the adversary attempts to recover the values of individual images, we obfuscate each sensitive sample by adding random noise to that sample. For each image, we randomly select 1/3 of pixels and add Gaussian noise $\mathcal{N}(0, 75)$ to the selected pixels. 

\begin{figure*}[t]
     \centering
     \subfloat[][Target images]{
     \includegraphics[width=\linewidth]{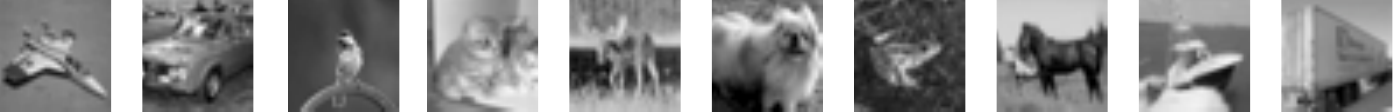}
     \label{fig:method:model_memorization_1}} \\
     \subfloat[][Recovered images]{
     \includegraphics[width=\linewidth]{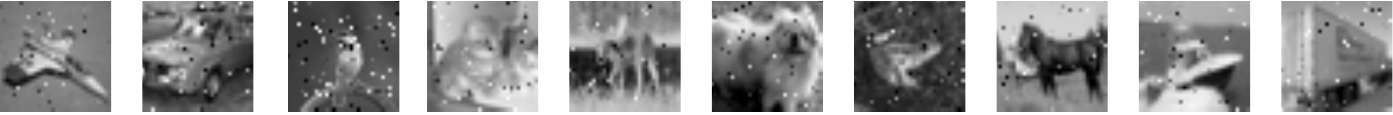}
     \label{fig:method:model_memorization_2}} \\
     \subfloat[][Recovered images with data obfuscation]{
     \includegraphics[width=\linewidth]{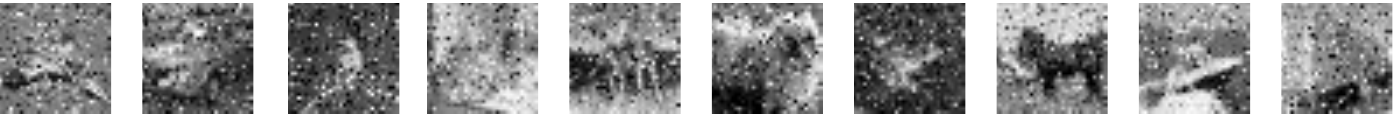}
     \label{fig:method:model_memorization_3}} 
\caption{Model memorization attack and the defense}
\label{fig:method:model_memorization}
\end{figure*}

\bheading{Privacy.}
We first consider the defense effects in privacy-preserving. Figure \ref{fig:method:model_memorization_3} shows the images the adversary recovers via model memorization technique. Since we add random noise to the images, the adversary only gets the noisy images. From these images, we observe that it is very difficult for the adversary to identify the objects and their labels. It is even harder to recognize the image details. So with data obfuscation, even the adversary has control of the training set, he cannot recover the clean and accurate data. 

\begin{figure}[t]
     \centering
     \subfloat[][Training accuracy]{
     \includegraphics[width=0.49\linewidth]{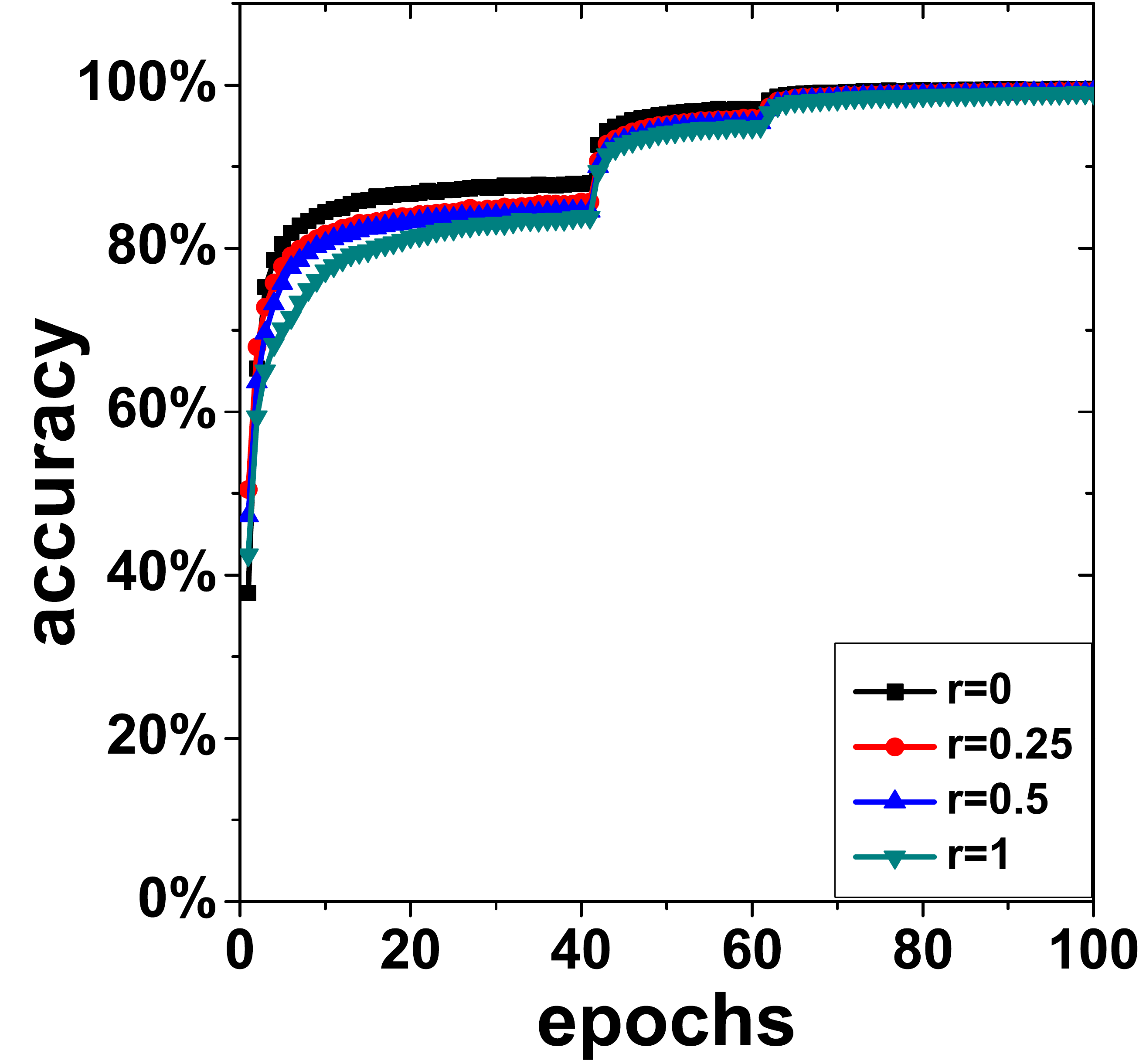}
     \label{fig:method:model_mem_training}} 
   \subfloat[][Validation accuracy]{
     \includegraphics[width=0.49\linewidth]{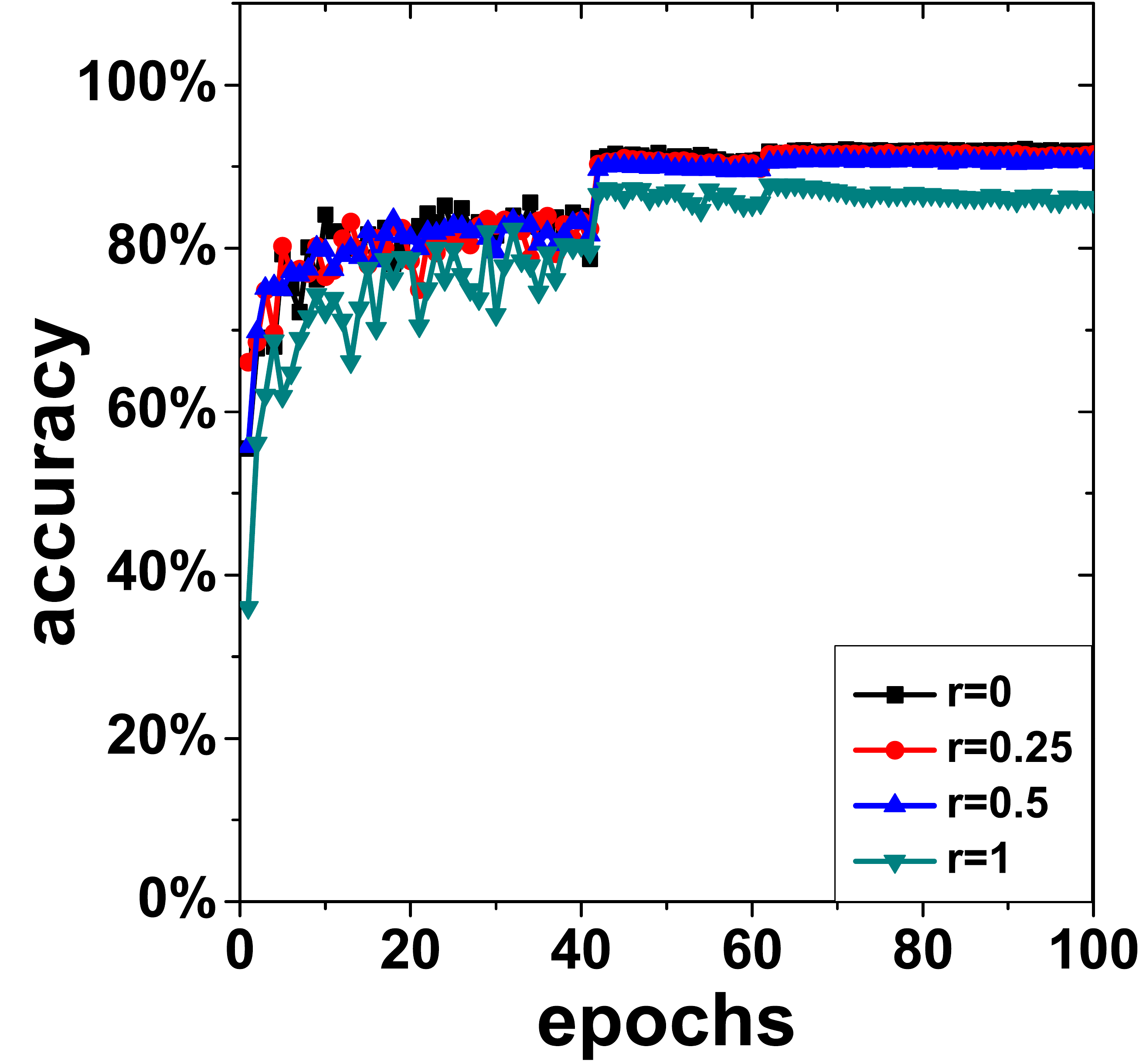}
     \label{fig:method:model_mem_testing}} 
    \caption{Prediction accuracy of machine learning models in model memorization attack.}
    \label{fig:model_mem:perf}
\end{figure}

\bheading{Accuracy.}
Next we consider the impact of adding noise to the model prediction accuracy. Figure \ref{fig:model_mem:perf} shows the training and validation accuracy versus training epochs. We consider different ratios of sensitive samples that are obfuscated ($r$ in Algorithm \ref{alg:ob_individual_sample}). We can see that the training time is not affected by the noise addition: each case reaches the maximum accuracy at almost the same time. However, there is a tradeoff between validation accuracy and noise ratio. When we add noise to 1/2 of the samples, the validation accuracy decreases 2\%. If we add noise to all of the samples, then the validation accuracy drops by 5\%, which is still acceptable. Noise addition is more effective when there are not so many sensitive samples to be protected. 

\subsection{Membership Inference Attack}

\subsubsection{Attack}

Next we consider membership inference attack \cite{ShStSo:17}, which also targets on the property of individual samples. For one specific sample $(x^*, y^*)$, the adversary wants to learn the membership property: $\mathcal{P}(x^*, y^*) = \mathbb{I}(x^*, y^* \in D)$, \ie, if the sample is used in the model training. The adversary does not have direct access to the training dataset or trained model. But he is able to use the trained model for prediction: he can send any feature data to the model and obtain the confidence score/probability that the input data belong to each class.

To infer the membership property, the adversary trains a lot of shadow models using the same algorithm of the target model. For each shadow model, the adversary randomly chooses some data samples to form a training set and a validation set. The shadow model is trained over the training set. Then he feeds both the training feature data and validation feature data into the shadow model and obtains the corresponding results, \ie, probabilities of each class. If the model is overfitted (this is an assumption in \cite{ShStSo:17}), the prediction results of training data and validation data are different. So the adversary can build a classifier, with the prediction results of feature data as input, and whether the data is from the training set (denoted as 1) or the validation set (denoted as 0) as the label. With this classifier, the adversary sends the feature value of sensitive sample in consideration to the target model and retrieves the prediction result. Then he sends the prediction result to the classifier and gets the conclusion whether the sensitive sample is a member of the target model's training set.

\bheading{Implementation.}
We reproduce one membership inference attack from \cite{ShStSo:17} with the same configuration and parameters: the target model is a convolutional neural network over the CIFAR10 dataset. We select two sizes of the training set for the target and shadow models: 2,500 and 10,000. The validation set has the same size as the training set. The adversary runs 100 shadow models with different subsets of training and validation samples. He uses logistic regression to classify whether the sample is included in the training set based on its prediction result. 

The black lines ($r=0$) in Figure \ref{fig:membership:acc} show the adversary's membership inference accuracy of samples in each class ($x$-axis), with 10,000 and 2,500 training samples respectively. We observe that the adversary can achieve high prediction accuracies for different classes ($80\% \sim 90\%$). Our reproduced results match the ones from \cite{ShStSo:17}. To better illustrate the membership inference attack effects, we show the confusion matrices in Tables \ref{table:r0n1} (10,000 training samples) and \ref{table:r0n0.25} (2,500 training samples). We can see that the adversary has high confidence of identifying the samples that are included in the training set. He has relatively high errors when the samples are not in the training set. In general the $F_1$ scores are very high (0.86 and 0.89). 

\subsubsection{Defense}
We use the same defense strategy of model memorization attack to prevent the adversary from inferring the membership of specific samples. We obfuscate each sensitive samples by injecting Gaussian noise. Then the target model is trained over noisy samples. When the adversary conducts the membership inference attack, his shadow models and the classifier are generated over the clean samples. If a sensitive sample is included in the target model's training set, the output of this sample from the target model will indicate that this sample is not in the training set, as the target model is trained with the noisy samples, not the clean ones. 

\begin{figure}[t]
     \centering
     \subfloat[][10k training samples]{
     \includegraphics[width=0.49\linewidth]{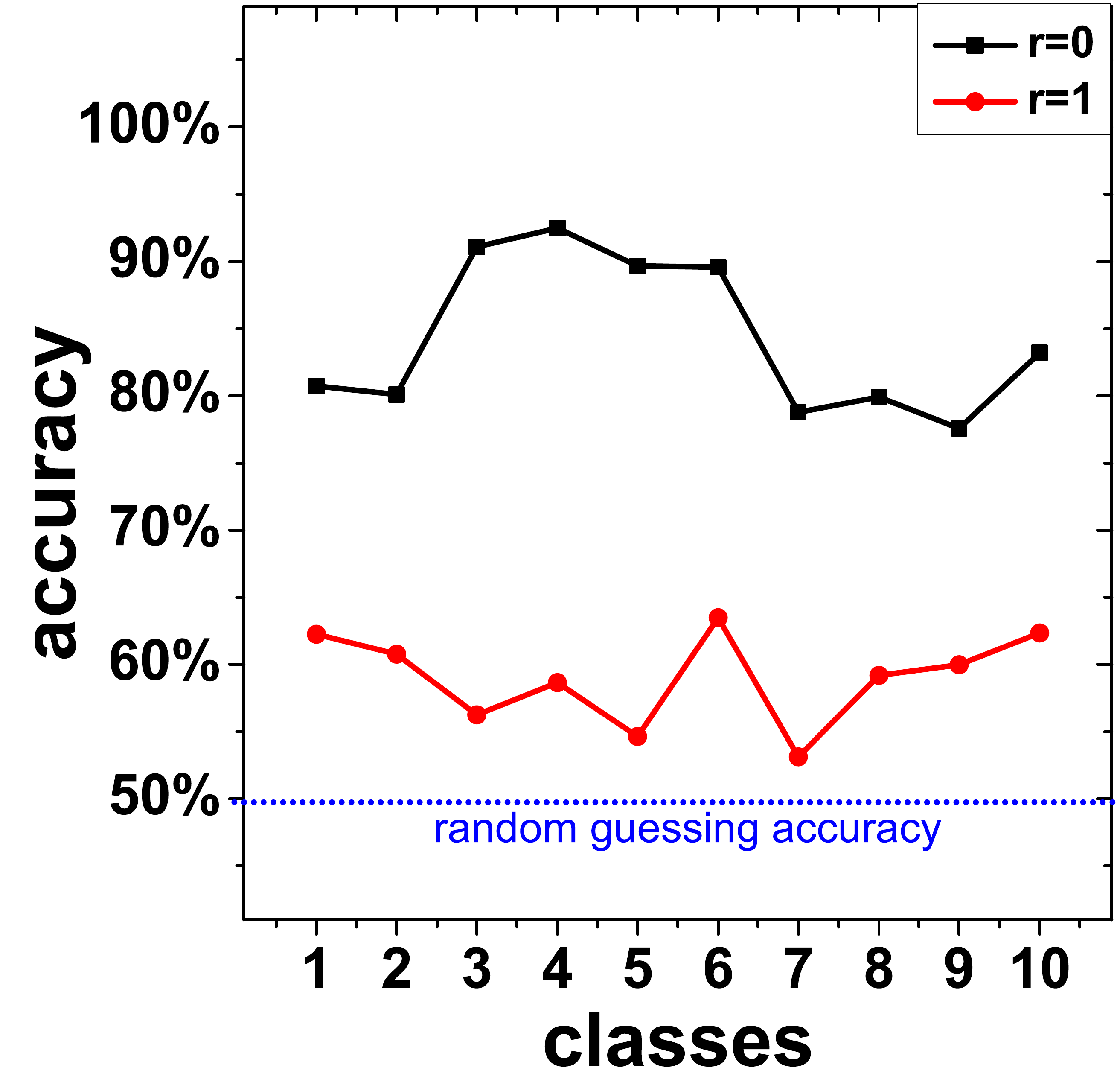}
     \label{fig:mi-acc-10k}} 
   \subfloat[][2.5k training samples]{
     \includegraphics[width=0.49\linewidth]{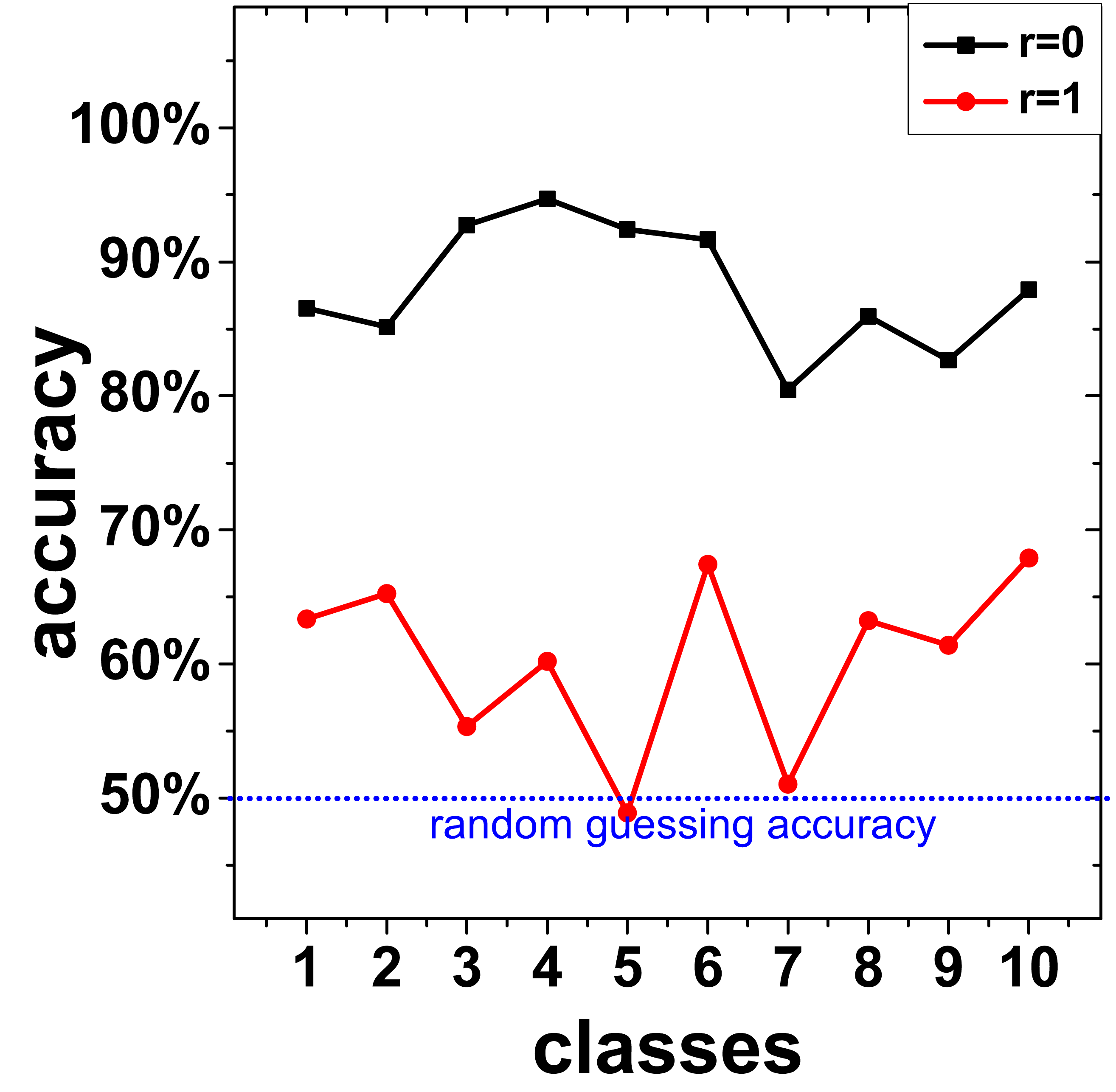}
     \label{fig:mi-acc-2.5k}} 
    \caption{Accuracy of the membership inference attack.}
    \label{fig:membership:acc}
\end{figure}

\bheading{Privacy.}
The red lines ($r=1$) in Figure \ref{fig:membership:acc} show the adversary's inference accuracy after the training samples are obfuscated. The base line is 50\% (blue dash line), which is the adversary's random guessing accuracy. With data obfuscation, the adversary's inference accuracy drops to $50\% \sim 65\%$, close to random guessing. This is much lower than the case without data obfuscation. Tables \ref{table:r1n1} and \ref{table:r1n0.25} shows the confusion matrices with the defense: the adversary's true positives are decreased significantly, as he will mis-classify the training samples as non-members of the training set.

\begin{figure}[t]
     \centering
     \subfloat[][Training accuracy]{
     \includegraphics[width=0.49\linewidth]{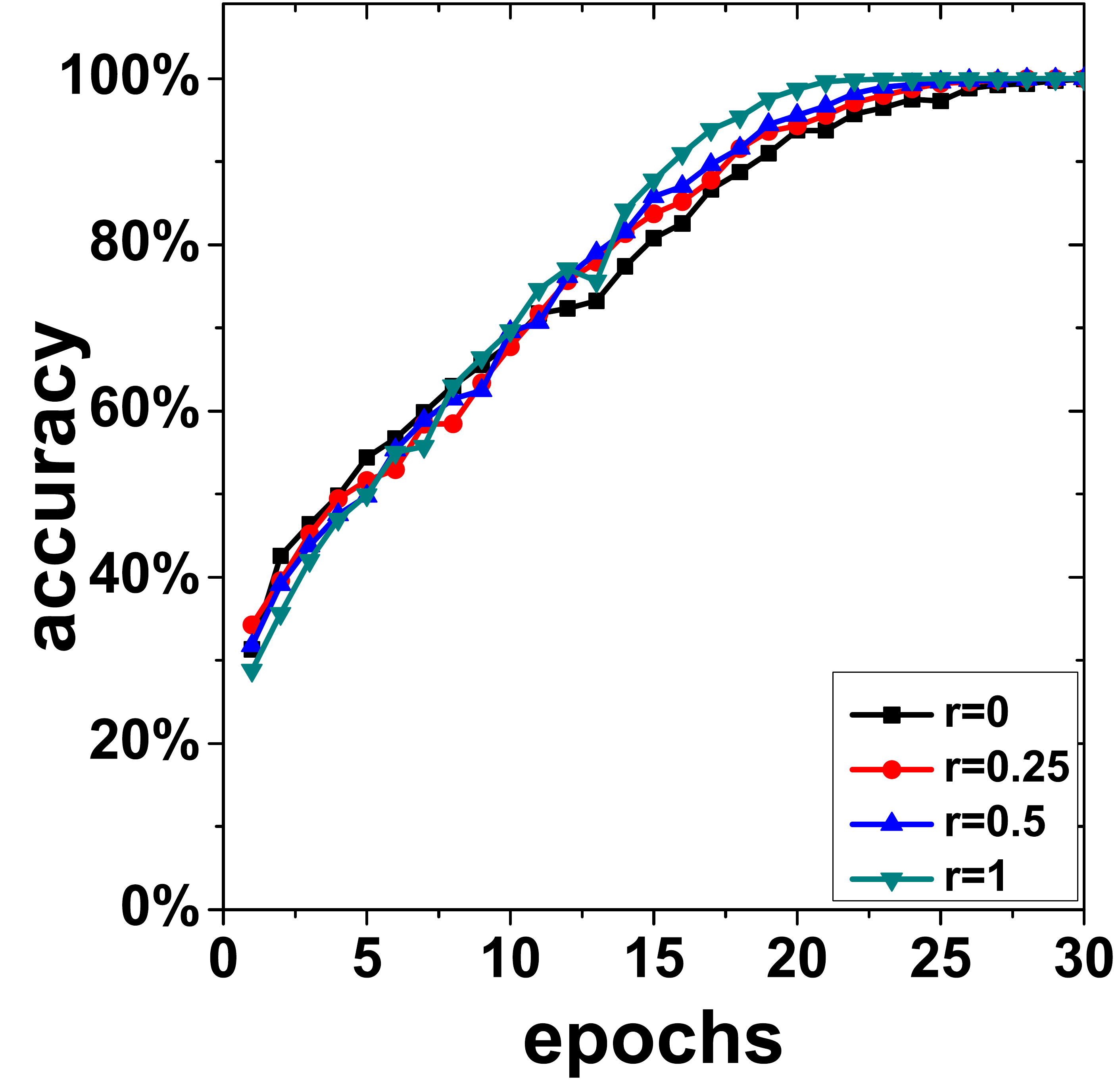}
     \label{fig:method:membership-train-10k}} 
   \subfloat[][Validation accuracy]{
     \includegraphics[width=0.49\linewidth]{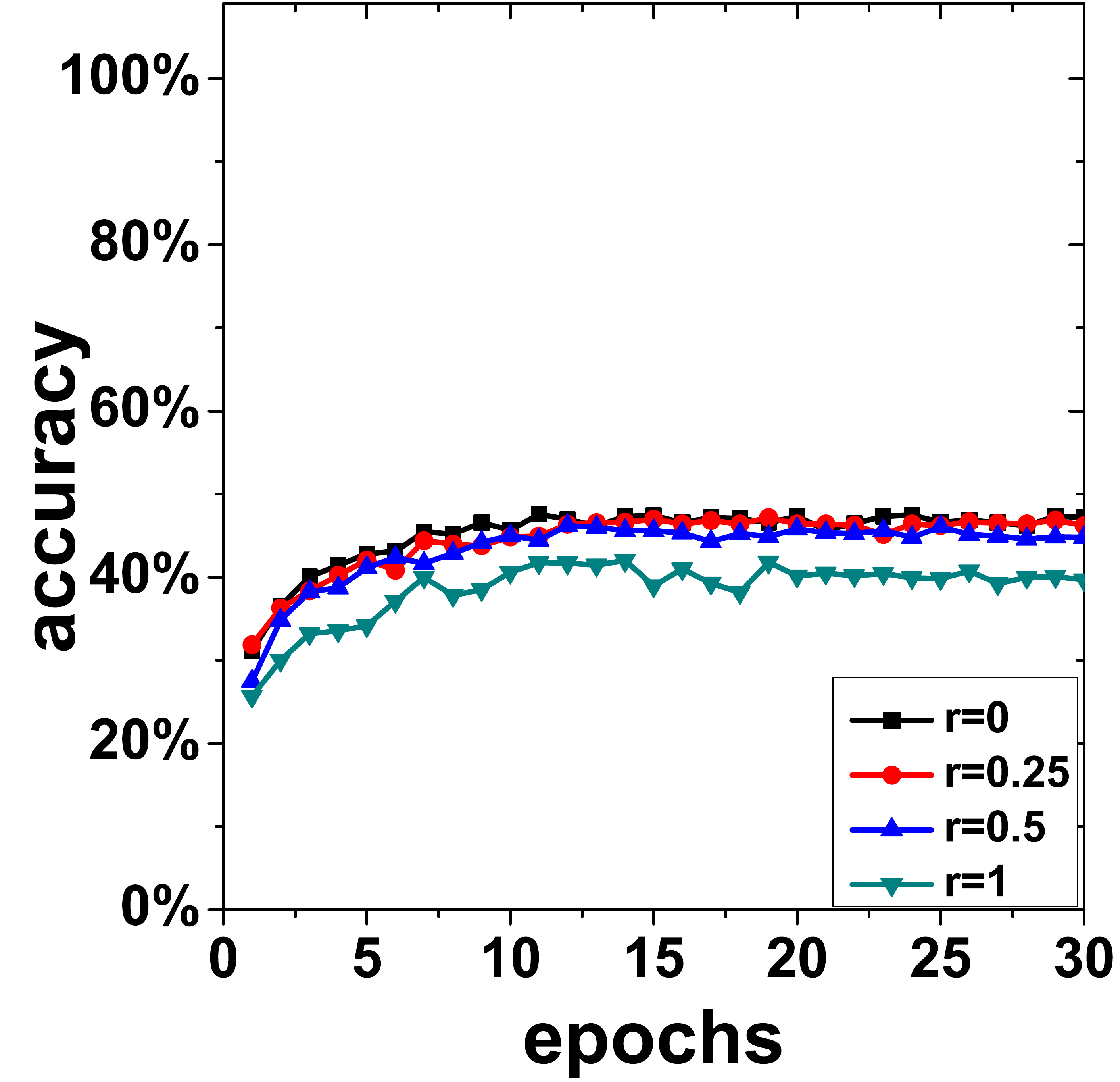}
     \label{fig:method:membership-test-10k.pdf}} 
    \caption{Prediction accuracy of machine learning models in membership inference attack (10k training samples).}
    \label{fig:membership:perf1}
\end{figure}

\begin{figure}[t]
     \centering
     \subfloat[][Training accuracy]{
     \includegraphics[width=0.49\linewidth]{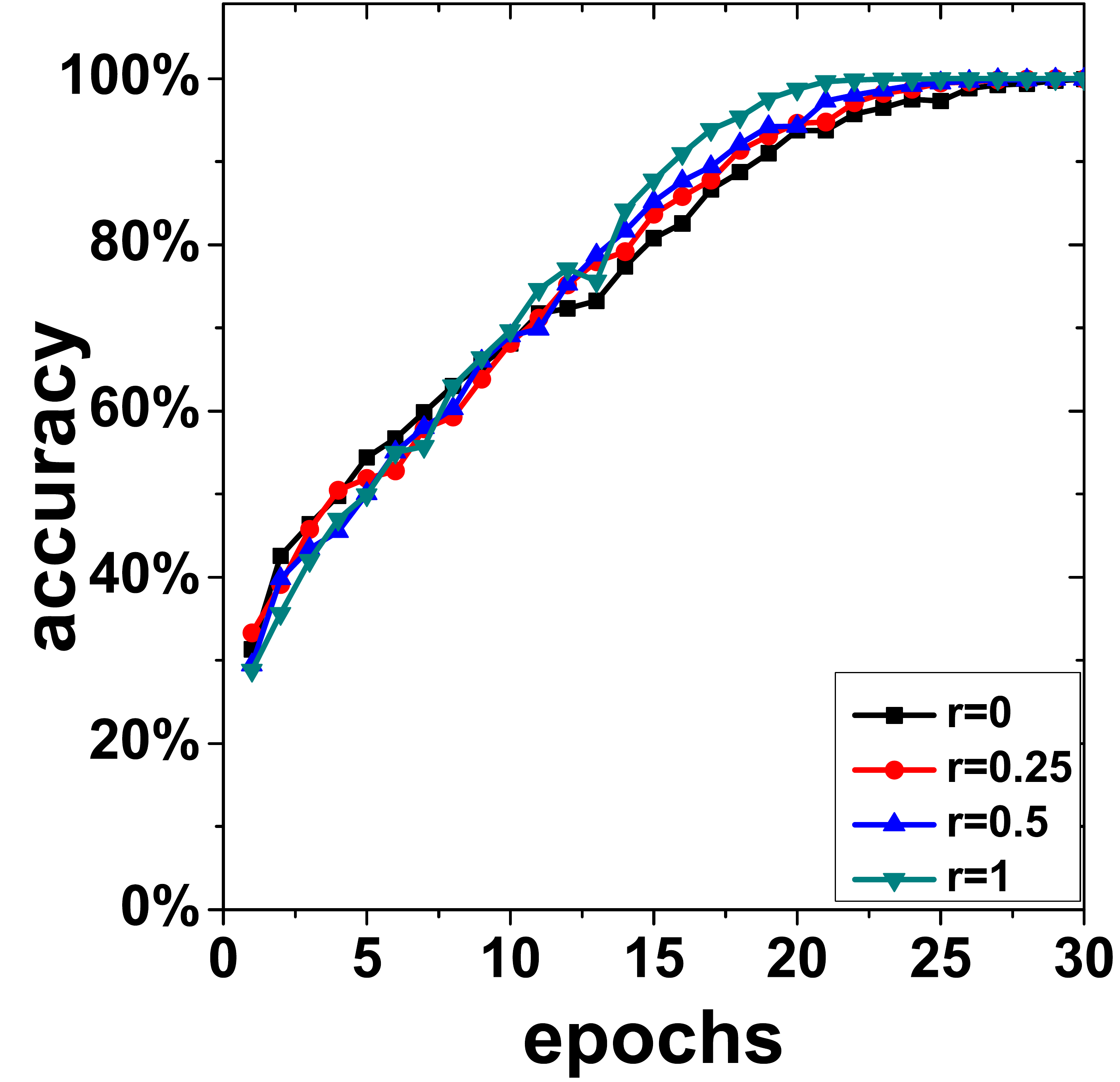}
     \label{fig:method:membership-train-25k}} 
   \subfloat[][Validation accuracy]{
     \includegraphics[width=0.49\linewidth]{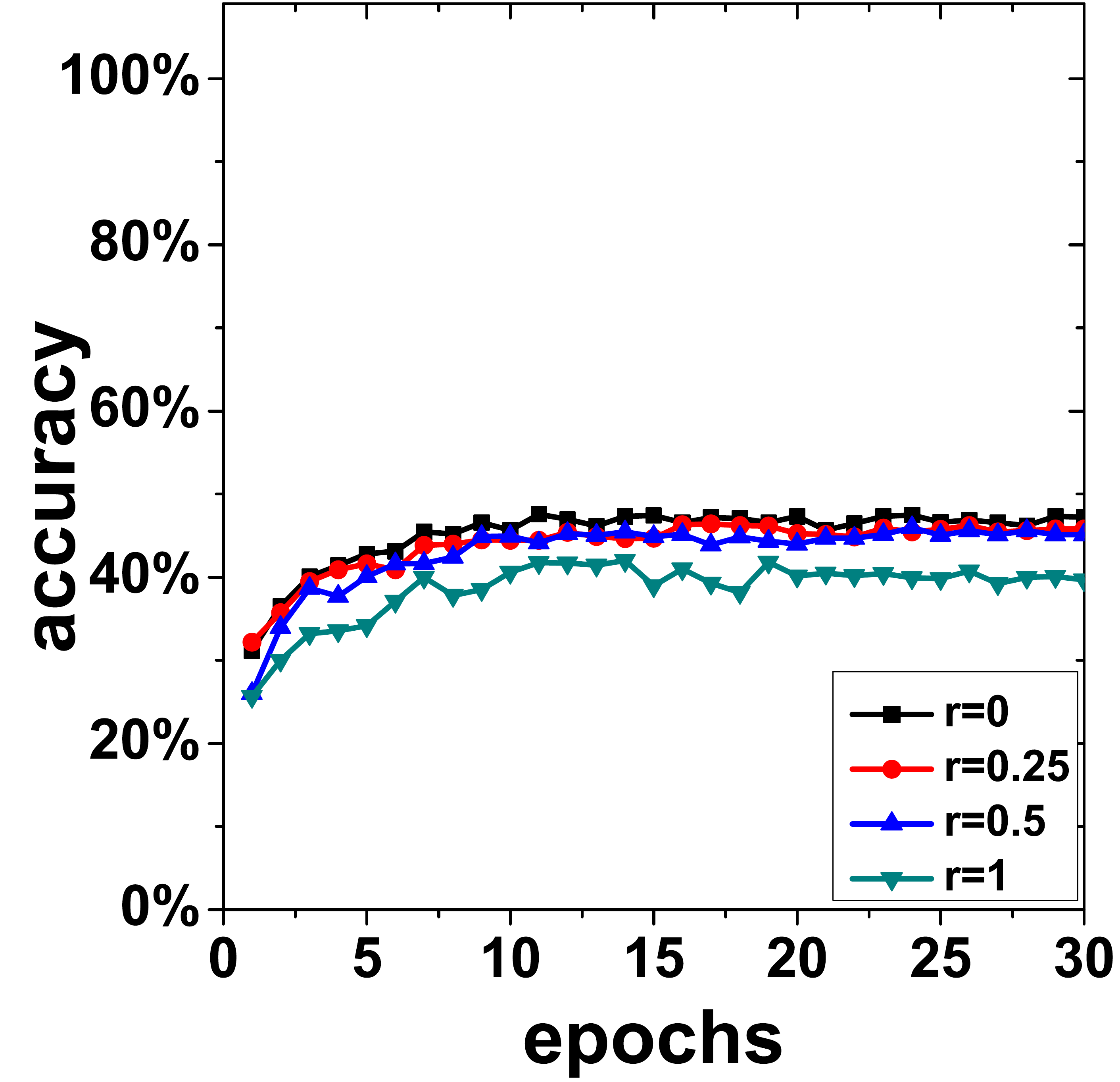}
     \label{fig:method:membership-test-25k.pdf}} 
    \caption{Prediction accuracy of machine learning models in membership inference attack (2.5k training samples).}
    \label{fig:membership:perf2}
\end{figure}

\bheading{Accuracy.}
We also need to consider the prediction accuracy of the target model with obfuscation. Figures \ref{fig:membership:perf1} and \ref{fig:membership:perf2} show the training and validation accuracy of the target model with two training set sizes. We consider different ratios ($r=0.25, 0.5, 1$) of sensitive samples that are obfuscated. First we observe that in membership inference attacks, the validation accuracy is much lower that the training accuracy, indicating that the target models are overfitted. Shokri et al. \cite{ShStSo:17} have the same conclusion and results, as the difference between training and validation accuracy gives the adversary the opportunity to judge if one sample is from training set or validation set. Second, obfuscating smaller ratios of samples will not affect the validation accuracy. When all the samples are obfuscated ($r=1$), the validation accuracy degradation is within 5\%, which is acceptable.

\begin{table}[ht]
\centering
\caption{Confusion matrix in membership inference attack (10k training samples)}
\label{table:mi_attack1}
\subfloat[][$r=0$: $F_1 = 0.86$]{
\label{table:r0n1}
\resizebox{4.2cm}{!}{
\begin{threeparttable}
\begin{tabular}{|c||c|c|}
  \hline
  &Predicted: & Predicted: \\
  & \textbf{in} & \textbf{out} \\
    \hline\hline
Actual: & \multirow{ 2}{*}{1} & \multirow{ 2}{*}{0} \\
\textbf{in} & & \\
  \hline
Actual: & \multirow{ 2}{*}{0.31} & \multirow{ 2}{*}{0.69} \\
\textbf{out} & & \\
  \hline
  \end{tabular}
  \end{threeparttable}
  }
  } 
\subfloat[][$r=1$: $F_1 = 0.52$]{
\label{table:r1n1}
\resizebox{4.2cm}{!}{
\begin{threeparttable}
\begin{tabular}{|c||c|c|}
  \hline
  &Predicted: & Predicted: \\
  & \textbf{in} & \textbf{out} \\
    \hline\hline
Actual: & \multirow{ 2}{*}{0.44} & \multirow{ 2}{*}{0.56} \\
\textbf{in} & & \\
  \hline
Actual: & \multirow{ 2}{*}{0.26} & \multirow{ 2}{*}{0.74} \\
\textbf{out} & & \\
  \hline
  \end{tabular}
  \end{threeparttable}
  }
  }
\end{table}

\begin{table}[ht]
\centering
\caption{Confusion matrix in membership inference attack (2.5k training samples)}
\label{table:mi_attack2}
\subfloat[][$r=0$: $F_1 = 0.89$]{
\label{table:r0n0.25}
\resizebox{4.2cm}{!}{
\begin{threeparttable}
\begin{tabular}{|c||c|c|}
  \hline
  &Predicted: & Predicted: \\
  & \textbf{in} & \textbf{out} \\
    \hline\hline
Actual: & \multirow{ 2}{*}{0.99} & \multirow{ 2}{*}{0.01} \\
\textbf{in} & & \\
  \hline
Actual: & \multirow{ 2}{*}{0.23} & \multirow{ 2}{*}{0.77} \\
\textbf{out} & & \\
  \hline
  \end{tabular}
  \end{threeparttable}
  }
  }  
\subfloat[][$r=1$: $F_1 = 0.51$]{
\label{table:r1n0.25}
\resizebox{4.2cm}{!}{
\begin{threeparttable}
\begin{tabular}{|c||c|c|}
  \hline
  &Predicted: & Predicted: \\
  & \textbf{in} & \textbf{out} \\
    \hline\hline
Actual: & \multirow{ 2}{*}{0.41} & \multirow{ 2}{*}{0.59} \\
\textbf{in} & & \\
  \hline
Actual: & \multirow{ 2}{*}{0.2} & \multirow{ 2}{*}{0.8} \\
\textbf{out} & & \\
  \hline
  \end{tabular}
  \end{threeparttable}
  }
  }  
\end{table}

\subsection{Model Inversion Attack}

\subsubsection{Attack}
Model inversion attack is a type of model privacy attack that can leak the statistical properties of samples with the same label. For instance, Fredrikson et al. \cite{FrJhRi:15} designed a model inversion attack, in which an adversary can recover recognizable images of people's faces. In this attack, the adversary is interested in a group $g^*$ of samples with the same feature, \ie, same label: $\mathcal{F}: y_{i} = y^{*}$. He wants to recover the statistical property of these samples, \ie, the average value: $\mathcal{P}(g^{*}) = avg (x_{i})$. The adversary retrieves the statistical property from the model parameters and outputs.

Recovering such statistical property is an optimization problem. For the above face recognition model inversion attack, given a person's name (label), the adversary's goal is to find the image, whose output from the model has the highest confidence score in that person's class. The adversary can use gradient descent method to solve this optimization problem and recover the image of that person.

\bheading{Implementation.}
We reproduce this attack from \cite{FrJhRi:15}. We adopt the AT\&T Face Database \cite{SaHaAd:96}. This database contains 40 classes with each one has 10 gray-scaled images of one person. We pick five persons from the database and their original faces are shown in Figure \ref{fig:method:model_inversion_1}. Figure \ref{fig:method:model_inversion_2} shows the corresponding images recovered by the adversary. We can see the recovered images are very close to the original ones and the adversary can easily identify the looking of one person from the attack results.

\subsubsection{Defense}
We use Algorithm \ref{alg:ob_group_sample} to hide the statistical property of sensitive samples. For the sensitive class, we add some new samples to the training set. These samples are synthesized as the negative values of the original images. So when the adversary conducts the model inversion attack to obtain the average value of the training samples in that class, the new samples will hide the information of the original ones.

\begin{figure}[t]
     \centering
     \subfloat[][Target images]{
     \includegraphics[width=\linewidth]{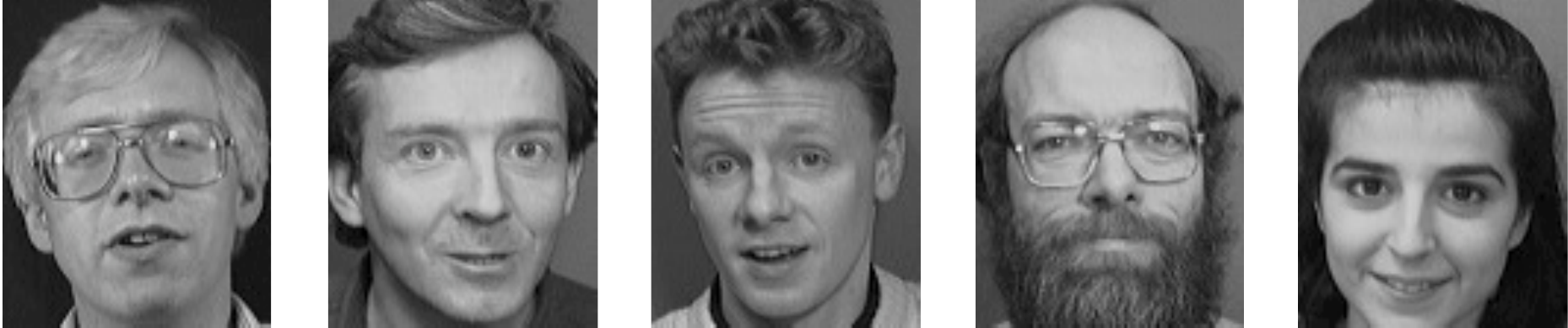}
     \label{fig:method:model_inversion_1}} \\
     \subfloat[][Recovered images]{
     \includegraphics[width=\linewidth]{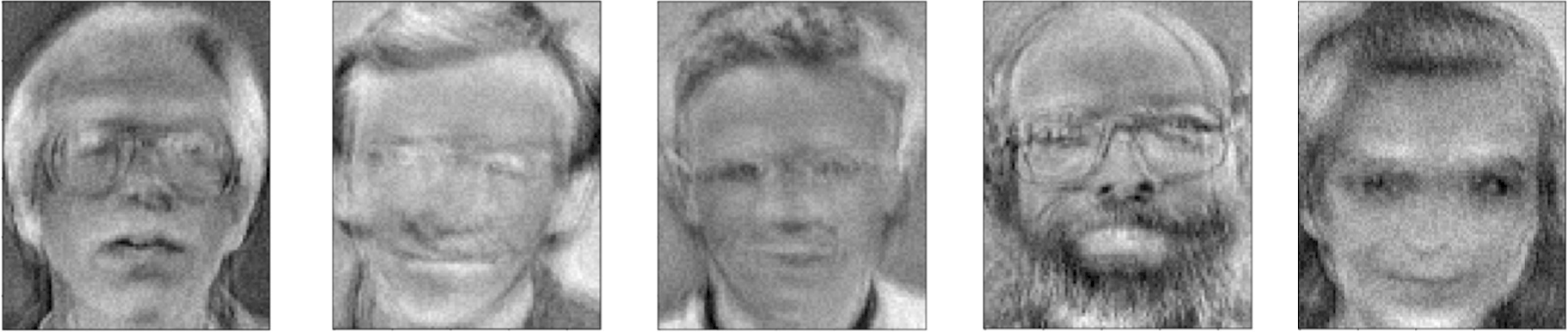}
     \label{fig:method:model_inversion_2}} \\
     \subfloat[][Recovered images with $r=0.25$]{
     \includegraphics[width=\linewidth]{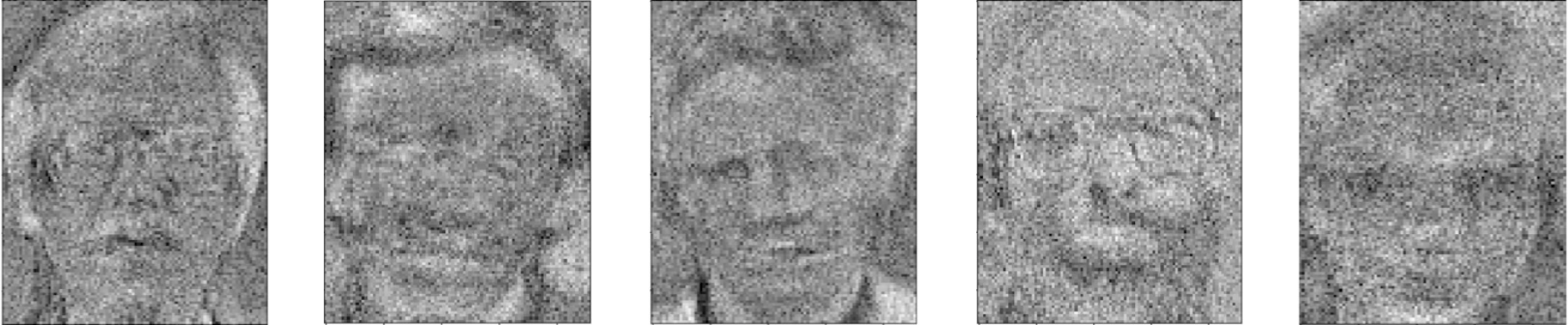}
     \label{fig:method:model_inversion_3}} \\
     \subfloat[][Recovered images with $r=0.5$]{
     \includegraphics[width=\linewidth]{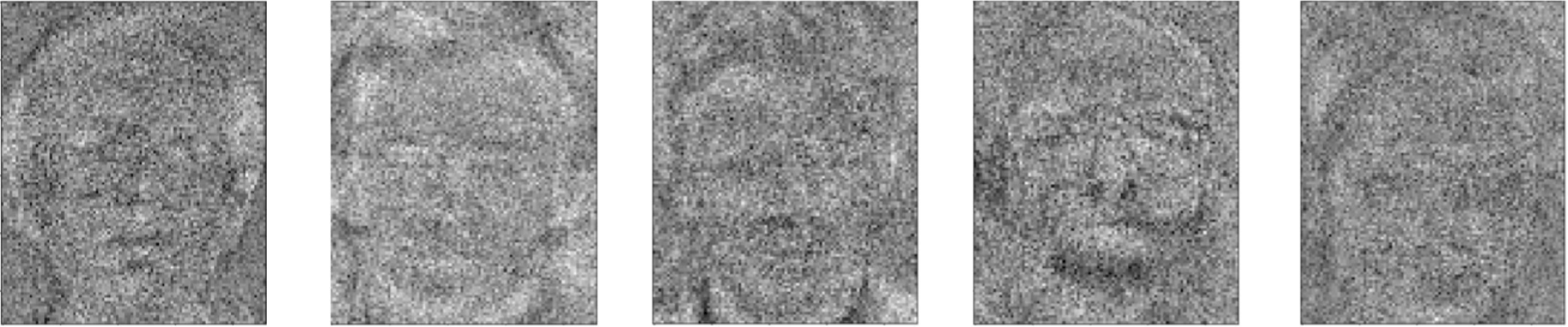}
     \label{fig:method:model_inversion_4}} \\
     \subfloat[][Recovered images with $r=1$]{
     \includegraphics[width=\linewidth]{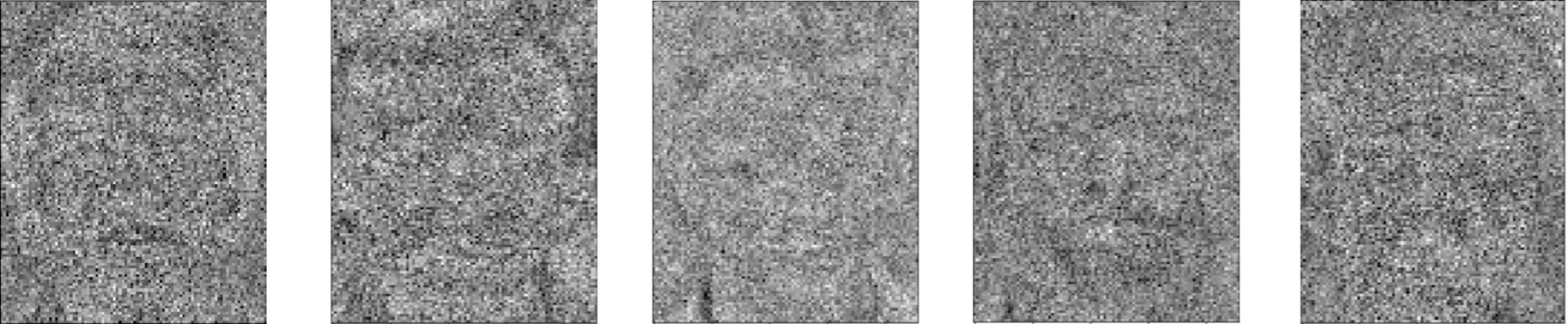}
     \label{fig:method:model_inversion_5}} \\
\caption{Model inversion attack and the defense}
\label{fig:method:model_inversion}
\end{figure}

\bheading{Privacy}.
Figures \ref{fig:method:model_inversion_3} -- \ref{fig:method:model_inversion_5} show the recovered images after we add different numbers of samples into the training set. We use $r$ to denote the ratio of the new samples to the original ones. We can observe that the added noisy samples can obscure the adversary's observations. This obfuscation operation is more effective when $r$ is larger. When we add the same number of new samples ($r=1$) to the training set, the adversary can hardly recognize the persons' faces.

\begin{figure}[t]
     \centering
     \subfloat[][Training accuracy]{
     \includegraphics[width=0.49\linewidth]{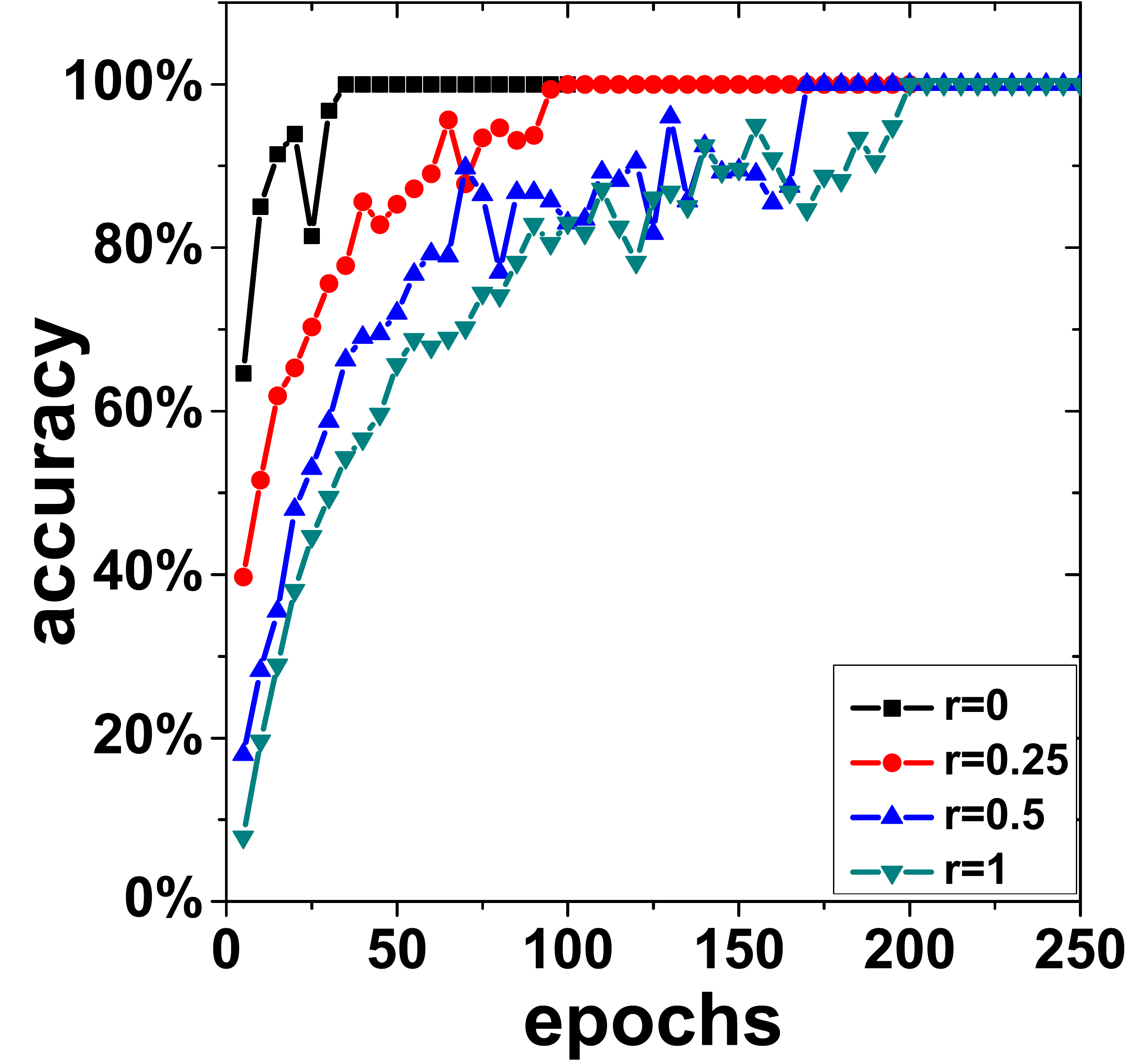}
     \label{fig:method:model_inversion_training}} 
   \subfloat[][Validation accuracy]{
     \includegraphics[width=0.49\linewidth]{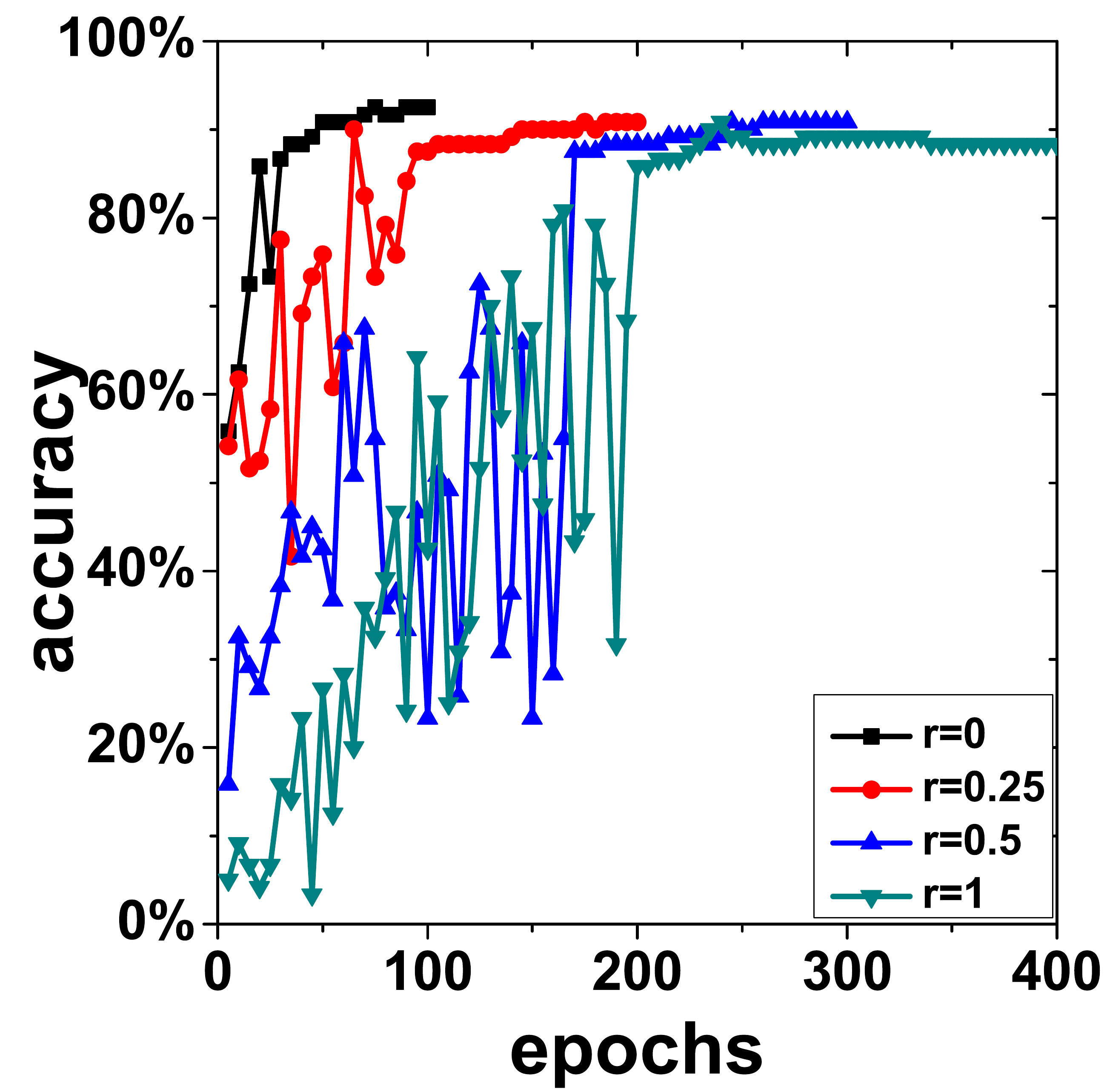}
     \label{fig:method:model_inversion_testing}} 
    \caption{Prediction accuracy of machine learning models in model inversion attack.}
    \label{fig:model_inversion:perf}
\end{figure}

\bheading{Accuracy.}
Figure \ref{fig:model_inversion:perf} shows the training and validation accuracies versus the number of training epochs with different ratios of new added samples. We use $r=0$ as the baseline. We can see that with the new synthetic samples, model training takes a little longer time to reach stable results. However, the model accuracy remains the same. This confirms that adding new samples do not affect the model accuracy of the old samples. 

\subsection{Model Classification Attack}

\subsubsection{Attack}
\label{sec:cc_attack}
Last, we consider model classification attacks that target on the entire training set ($\mathcal{F}: x_{i}, y_{i} \in D$) \cite{AtMaSp:15}. The adversary wants to learn some attributes of the training set, $\mathcal{P}(g^*=D) = attr(x_i)$. Ateniese et al. \cite{AtMaSp:15} gave two examples about the training set attributes: inferring if the voice training samples contain Indian accents in a speech recognition, and inferring if Google web traffic is used in a network traffic classification task. In this attack setting, the adversary is assumed to have access to the model parameters, and infer the attributes from these parameters.

Similar to membership inference attack, the adversary can build a classifier to predict the property. Specifically, the adversary can first train a number of shadow models. Some of the shadow models are trained over the training set with the property $\mathcal{P}$. These shadow models are labeled as 1. The rest shadow models are trained over the training set without the property. These shadow models are labeled as 0. Then the adversary retrieves the parameters of each shadow model as the features, and trains a new classifier over these features and model labels. This classifier can map the model parameters to the model labels, \ie,  whether the training set used in this model has the property or not.  So given the target model, the adversary can retrieve its model parameters, feed them into the classifier and predict if the training set of this model has the property.

\begin{table*}[ht]
\centering
\caption{Confusion matrix in model classification attack}
\label{table:cc_attack}
\subfloat[][$r=0$: $F_1 = 0.995$]{
\label{table:r0}
\resizebox{4.2cm}{!}{
\begin{threeparttable}
\begin{tabular}{|c||c|c|}
  \hline
  &Predicted: & Predicted: \\
  & \textbf{MNIST} & \textbf{non-MNIST} \\
    \hline\hline
Actual: & \multirow{ 2}{*}{0.99} & \multirow{ 2}{*}{0.01} \\
\textbf{MNIST} & & \\
  \hline
Actual: & \multirow{ 2}{*}{0} & \multirow{ 2}{*}{1} \\
\textbf{non-MNIST} & & \\
  \hline
  \end{tabular}
  \end{threeparttable}
  }
  } 
\subfloat[][$r=0.25$: $F_1 = 0.246$]{
\label{table:r0.25}
\resizebox{4.2cm}{!}{
\begin{threeparttable}
\begin{tabular}{|c||c|c|}
  \hline
  &Predicted: & Predicted: \\
  & \textbf{MNIST} & \textbf{non-MNIST} \\
    \hline\hline
Actual: & \multirow{ 2}{*}{0.14} & \multirow{ 2}{*}{0.86} \\
\textbf{MNIST} & & \\
  \hline
Actual: & \multirow{ 2}{*}{0} & \multirow{ 2}{*}{1} \\
\textbf{non-MNIST} & & \\
  \hline
  \end{tabular}
  \end{threeparttable}
  }
  }
\subfloat[][$r=0.5$: $F_1 = 0.198$]{
\label{table:r0.5}
\resizebox{4.2cm}{!}{
\begin{threeparttable}
\begin{tabular}{|c||c|c|}
  \hline
  &Predicted: & Predicted: \\
  & \textbf{MNIST} & \textbf{non-MNIST} \\
    \hline\hline
Actual: & \multirow{ 2}{*}{0.11} & \multirow{ 2}{*}{0.89} \\
\textbf{MNIST} & & \\
  \hline
Actual: & \multirow{ 2}{*}{0} & \multirow{ 2}{*}{1} \\
\textbf{non-MNIST} & & \\
  \hline
  \end{tabular}
  \end{threeparttable}
  }
  }  
\subfloat[][$r=1$: $F_1 = 0.077$]{
\label{table:r1}
\resizebox{4.2cm}{!}{
\begin{threeparttable}
\begin{tabular}{|c||c|c|}
  \hline
  &Predicted: & Predicted: \\
  & \textbf{MNIST} & \textbf{non-MNIST} \\
    \hline\hline
Actual: & \multirow{ 2}{*}{0.04} & \multirow{ 2}{*}{0.96} \\
\textbf{MNIST} & & \\
  \hline
Actual: & \multirow{ 2}{*}{0} & \multirow{ 2}{*}{1} \\
\textbf{non-MNIST} & & \\
  \hline
  \end{tabular}
  \end{threeparttable}
  }
  }  
\end{table*}

\bheading{Implementation.}
Since detailed attack code, machine learning algorithms and training dataset are not opensourced in \cite{AtMaSp:15}, we implement a new attack using the same technique. 
We consider a handwritten digit classification task, which uses LeNet neural network to classify digits ``0'' -- ``9'' from handwritten images. The adversary wants to learn the property that \emph{if the training set is MNIST \cite{LeBoBe:98} or not}. In our experiment, the adversary trains 100 shadow models with MNIST and 100 shadow models with another handwritten image dataset, EMNIST \cite{CoAfTa:17}. For each shadow model, the adversary retrieves all the trainable parameters in each neural network layer (\texttt{conv1/kernel}, \texttt{conv1/bias}, \texttt{conv2/kernel}, \texttt{conv2/bias}, \texttt{fc1/kernel}, \texttt{fc1/bias}, \texttt{fc2/kernel}, \texttt{fc2/bias}), and calculates the mean and standard deviation of each parameter vector. This generates 16 values, representing the feature of the shadow model. Then the adversary builds a classifier using logistic regression to classify the MNIST or non-MNIST shadow models from the feature values. To measure the adversary's classification accuracy, we train another 100 models with MNIST and 100 models with EMNIST for validation. Table \ref{table:r0} shows the confusion matrix of the classification. We can observe that the classifier has a very good performance: only 1 model with non-MNIST is mis-classified as EMNIST. The F1-score is 0.995.

\subsubsection{Defense}
Similar to model inversion attack, we can obfuscate the entire training data by augmenting the dataset with synthetic samples that are negative of original ones (Algorithm \ref{alg:ob_group_sample}). With these faked samples, the statistical property of the original dataset is hidden from the adversary's classifier.

\bheading{Privacy.}
Tables \ref{table:r0.25} -- \ref{table:r1} show the confusion matrices of the adversary's classifier when we add different ratios of synthetic samples. We can observe that with the augmented dataset, most MNIST models will be classified as non-MNIST, and the defense is more effective when the added sample ratio $r$ increases. One interesting observation is that adding new samples cannot make the adversary mis-classify the non-MNIST model as MNIST model. This is because EMNIST is already an extended version of MNIST and contains different types of handwritten digits. So adding noise to this dataset makes it more extended and less like pure MNIST data. But this is already effective in protecting the statistical properties of training set: when the adversary feeds the target model into his classifier, he will always get the non-MNIST label, and cannot tell if the training dataset of the target model is MNIST or not. 

\begin{figure}[t]
     \centering
     \subfloat[][Training accuracy]{
     \includegraphics[width=0.49\linewidth]{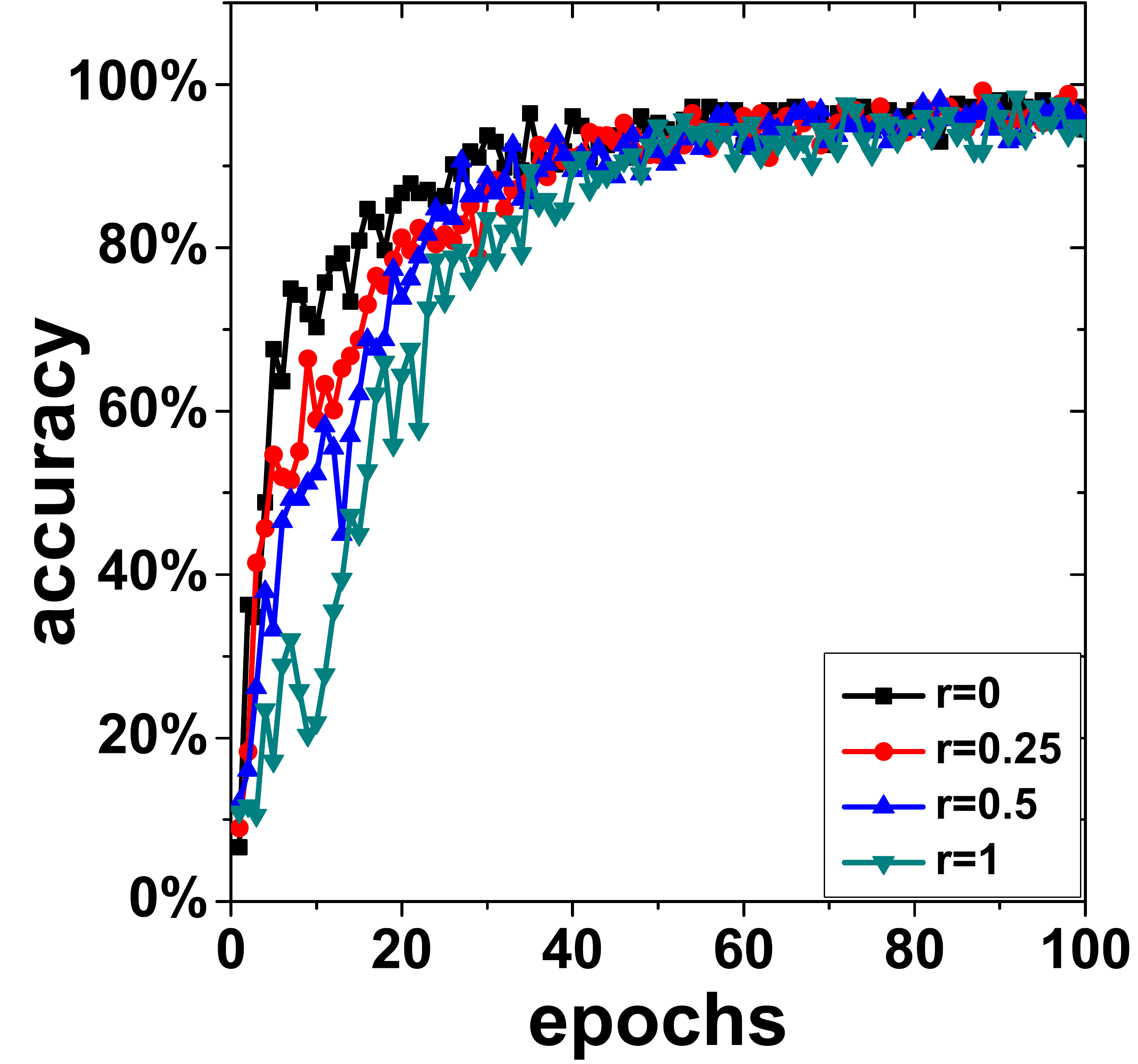}
     \label{fig:method:class_class_training}} 
   \subfloat[][Validation accuracy]{
     \includegraphics[width=0.49\linewidth]{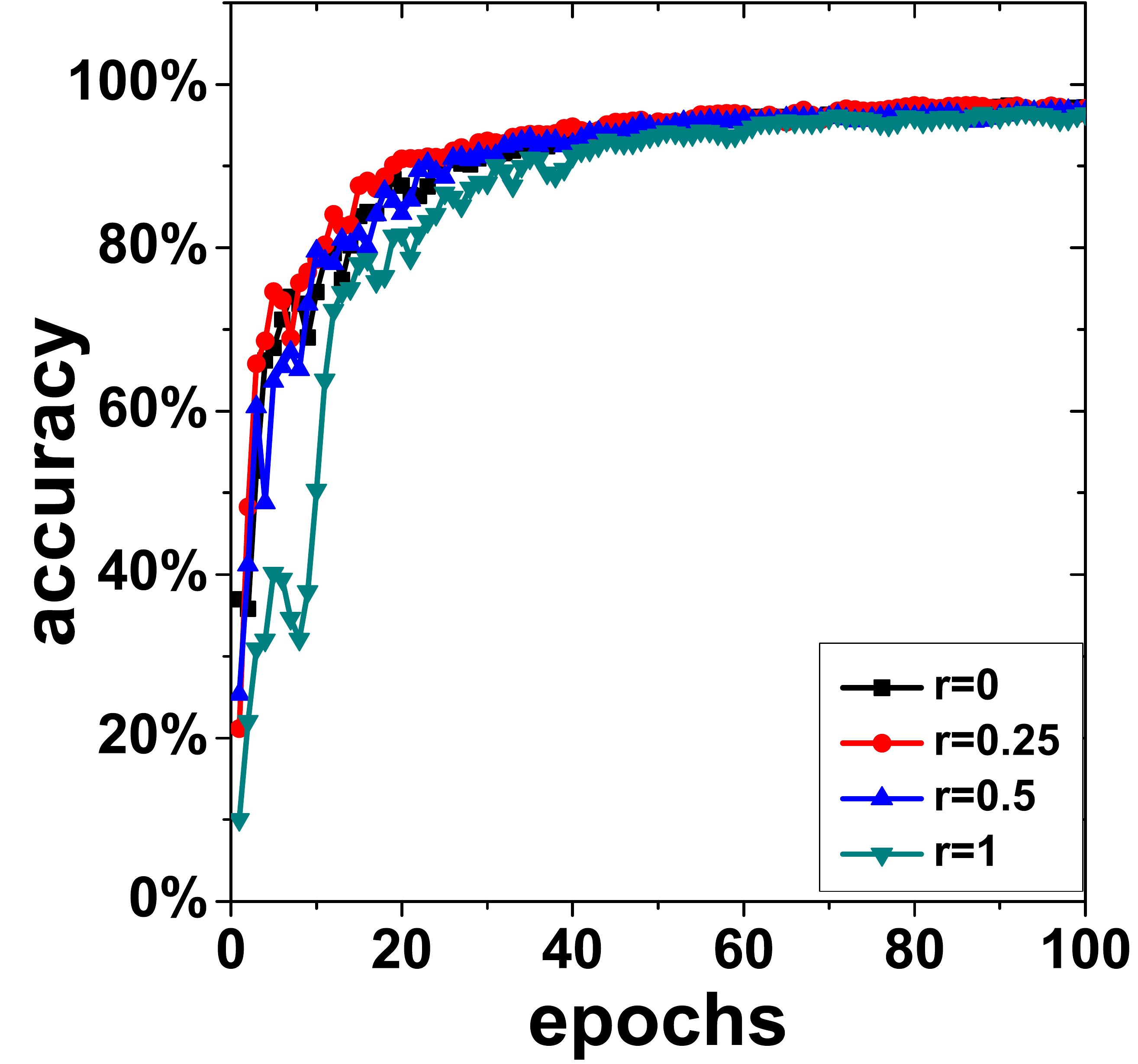}
     \label{fig:method:class_class_testing}} 
    \caption{Prediction accuracy of machine learning models in model classification attack.}
    \label{fig:class_class:perf}
\end{figure}

\bheading{Accuracy.}
Figure \ref{fig:class_class:perf} shows the prediction accuracy of the training set that is augmented and the validation set during the training process. We observe that adding new noisy samples to the training set has very small impact to the training performance: when adding the same number of synthetic samples as the original ones, it takes almost the same amount of time to reach the same high accuracy. 

\section{Related Work}
\label{sec:related}

\subsection{Machine Learning Privacy Attacks}
\label{sec:related:attack}

Different types of attacks were designed to breach the privacy of machine learning models or datasets. 

\bheading{Training data privacy.}
Training data is a popular target for adversaries to steal during the model training or serving phases. Ateniese et al. \cite{AtMaSp:15} proposed the model classification attack: given a target model, an adversary is able to extract some meaningful attributes of the training dataset. The basic attack strategy is to train different shadow models over datasets with different attributes. Then the adversary can build a new classifier which can predict the dataset attributes based on a model parameter as the feature. 

Shokri et al. \cite{ShStSo:17} proposed the membership inference attacks to infer if one individual sample is included in the training set of the target model. The attack technique is similar to \cite{AtMaSp:15}: the adversary builds a quantity of shadow models with different training datasets. He collects the prediction results of the training samples and validation samples from each shadow model. Using the prediction results as features and whether the sample is from training set or validation set as the labels, the adversary can build a new classifier to predict the membership given a sensitive sample and a target model. Salem et al. \cite{SaZhHu:18} simplified the attacks by relaxing some assumptions, \eg, using multiple shadow models, knowledge of the target model structure and dataset from the same distribution. Further studies show that membership inference attacks are possible even when the target models are not overfitted\cite{YeGiFr:18, LoBiWa:18}. Membership inference attacks are also demonstrated in Generative Adversarial Networks \cite{JaLuGe:17, LiLiGa:18}. Melis et al. \cite{MeSoDe:18} studies the inference attacks in collaborative learning: the adversary can learn the membership of training data used by other participants, based on the shared parameters in a distributed learning system. 

Fredrikson et al. \cite{FrLaJh:14} proposed the model inversion attacks: given a machine learning model, part of a sensitive sample's features and its label, the adversary is able to recover the rest features of this sample. The adversary's strategy is to reverse the model training process: figuring out the mapping from the output to input and predicting the input feature that best matches the output label. An advanced model inversion attack over deep neural network was proposed in \cite{FrJhRi:15} which is able to inverse an image classification model and recover an image given the corresponding label of the image. Hitaj et al. \cite{HiAtPe:17} demonstrated the possibility of model inversion attacks in collaborative learning even under privacy-preserving protection. They adopted the Generative Adversary Network to actively recover the private data samples used by other participants. 

Song et al. \cite{SoRiSh:17} proposed model memorization attacks, in which the adversary has direct access to the training data. He attempts to encode the sensitive data into the model for a receiver entity to retrieve them. Two cases were considered: if the receiver entity has white-box access to the produced model, then the adversary can directly encode the training data into the model parameters; if the receiver party only has black-box access to the model, then the adversary can encode the training data into the outputs corresponded to some specific inputs. In these ways, the private training samples are leaked from model parameters or outputs.

\bheading{Model privacy.}
In addition to the training set, adversaries may also be interested in the privacy of machine learning models. Tram{\`e}r et al. \cite{TrZhJu:16} proposed an attack to steal machine learning models using prediction APIs. The basic idea is that the adversary can generate a quantity of input samples and feed them into the target model for prediction. With the predicted outputs, the adversary can discover the model that best represents the relationships between inputs and outputs. Oh et al. \cite{OhAuFr:18} built metamodels to extract more model details such as neural network architecture, which is assumed public in \cite{TrZhJu:16}. Wang et al. \cite{WaGo:18} designed attacks to steal the hyperparameter of the machine learning model. A hyperparameter is used to balance the loss function and regularization term in the objective function. The adversary can obtain this value from the training set and model. The hyperparameter can further help the adversary recover the model parameters in a more efficient way. 

In this paper we consider the privacy attacks on the training data, so model privacy is out of the scope, as models are usually public and less private than the training dataset. We design a generic methodology that can cover different types of privacy attacks with different adversary's capabilities, and protect the privacy of different properties of training data. 

\subsection{Machine Learning Privacy Solutions}
\label{sec:related:defense}

\bheading{Enhancing the algorithms.}
One solution is to design secure algorithms to prevent information leakage during model training. Shokri and Shmatikov \cite{ShSh:15} proposed to use distributed training to protect the privacy of training data: different participants can train the model with their data locally and only upload the parameters with differential privacy noise, so training data will not be leaked to unauthorized parties. Hamm et al. \cite{HaChCh:15} applied the same idea to privacy machine learning in distributed IoT sensing systems. Cao et al. \cite{CaYa:15} proposed a methodology to modify the training algorithms to remove the effects of sensitive training samples on the models. Ohrimenko et al. \cite{OhScFo:16} proposed to use Intel SGX-processors to protect the training tasks against privileged adversaries in the host environment. They also designed data-oblivious machine learning algorithms that can eliminate the data-dependent accesses. Similarly, Hunt et al. \cite{HuSoSh:18} used Intel SGX to restrict the confidential training data from being leaked to adversaries. Abadi et al. \cite{AbChGo:16} applied differential privacy mechanism to design privacy-preserving deep learning algorithm: they added noise during the stochastic gradient descent process to eliminate the model parameters' dependency on the training dataset. 

The above methodologies have some drawbacks. First, the distributed privacy-preserving approach \cite{ShSh:15} can only protect the privacy of individual samples, but does not prevent property inference attacks that apply to groups of training samples \cite{MeSoDe:18}. Second, for the above methods, we must assume that the designed machine learning algorithms and schemes can be correctly deployed in MLaaS. However, if the adversary takes control of the machine learning platform, he may not run the enhanced algorithms or systems for privacy protection. Third, if the adversary has direct access to the training data \cite{SoRiSh:17}, then enhancing the training process does not work. Fourth, for SGX-assistant solutions, new hardware support is required, which makes it hard and impractical to deploy. Besides, SGX solutions cannot be applied to GPU architecture, which is now widely used in machine learning tasks.

\bheading{Enhancing the dataset.}
Enhancing the dataset has advantages over enhancing the training algorithms: even the adversary has direct access to the training data or MLaaS platform, he cannot get the sensitive data from the enhanced set. Bost et al. \cite{BoPoTu:15} proposed to encrypt the dataset before feeding them into the training algorithm. They designed machine learning operators which can operate on the encrypted data. By doing so the adversary cannot get the data if he does not have the encryption key. Although Homomorphic encryption works but it is hard to be applied to any arbitrary algorithm, and the performance is not efficient. 

Triastcyn and Faltings \cite{TrFa:18} proposed to use Generative Adversarial Network with differential-privacy to generate artificial data for deep learning. This idea is similar to our data obfuscation methodology. However, their approach is lack of strict differential privacy guarantees. The artificial data generated by their approach is quite close to real samples, with very minor differences in details. This is not enough for privacy preservation. Besides, they cannot defeat inference attacks against statistical properties, as the new samples still keep the characteristics of original ones. In contrast, we exploit the obfuscation operations on the training dataset to hide the sensitive properties while maintaining the model's prediction accuracy. We empirically show that this solution can defeat different types of data privacy attacks.

\section{Conclusion}
\label{sec:conclu}

The fast development of machine learning techniques have raised the security concern of data privacy. Past work have shown that an adversary can easily steal properties about the sensitive training samples in a MLaaS platform. 
We propose a generic approach to protect the privacy of training data by obfuscating the training set. We consider the protection of sensitive properties of individual samples, as well as statistical properties of groups of samples. Our empirical evaluations indicate this method can effective defeat a wide range of existing privacy attacks without affecting the training performance and prediction accuracy. Future work include the optimization of adding noise to individual samples by selecting minimal key feature values, and using GAN to generate new dataset to hide statistical properties.

\bibliographystyle{abbrv}
\bibliography{ref}

\end{document}